\documentclass[twocolumn,twoside]{IEEEtran}

\def\home {macros}%{C:/Users/anima/Documents/research/ltxtemplate}
\def\fighome {figures}%{C:/Users/anima/Documents/research/det_graphs/AnandkumarTongSwami07ICASSP/figures}

\usepackage[cmex10]{amsmath}
\usepackage{epsfig,epsf,psfrag,amssymb,amsfonts,latexsym,slashbox,graphicx,bm,cite,xcolor,url}
\usepackage[caption=false]{subfig}%,font=footnotesize
\usepackage{fixltx2e}
\usepackage{array}
\usepackage{cases}%[subnum]
\usepackage{verbatim}
\usepackage[mathscr]{eucal}
\usepackage{algpseudocode}%

\input \home/atbeginend.sty%
\def\plim{\mbox{p\,}\lim}

\newcommand\indep{\protect\mathpalette{\protect\independenT}{\perp}}
\def\independenT#1#2{\mathrel{\rlap{$#1#2$}\mkern2mu{#1#2}}}

\def\viz{{viz.,\ \/}}
\def\ie{{i.e.,\ \/}}
\def\eg{{e.g.,\ \/}}

\def\as{{a.s.  }}

\def\nn{\nonumber}

\def\qed{\hfill\hbox{${\vcenter{\vbox{
    \hrule height 0.4pt\hbox{\vrule width 0.4pt height 6pt
    \kern5pt\vrule width 0.4pt}\hrule height 0.4pt}}}$}}

\definecolor{myred}{rgb}{0.3,0.0,0.7}
\definecolor{dkg}{rgb}{0.1,0.7,0.2}
\definecolor{dkb}{rgb}{0.0,0.2,0.8}

\newcommand{\Cmsc}{\mathscr{C}}

\newcommand{\Emsc}{\mathscr{E}}
\newcommand{\Fmsc}{\mathscr{F}}
\newcommand{\Gmsc}{\mathscr{G}}

\def\bfv{{\mathbf v}}

\def\bfy{{\mathbf y}}
 
\def\bfV{{\mathbf V}}

\def\bfY{{\mathbf Y}}
    
\def\Bc{{\cal B}}
\def\Cc{{\cal C}}

\def\Ec{{\cal E}}

\def\Hc{{\cal H}}

\def\Pc{{\cal P}}

\def\Xc{{\cal X}}

\def\Ebb{{\mathbb E}}

\newcommand{\bprf}{\begin{myproof}}
\newcommand{\eprf}{\end{myproof}}
\newcommand{\bp}{\begin{psfrags}}
\newcommand{\ep}{\end{psfrags}}
\newcommand{\bl}{\begin{lemma}}
\newcommand{\el}{\end{lemma}}
\newcommand{\bt}{\begin{theorem}}
\newcommand{\et}{\end{theorem}}
\newcommand{\bc}{\begin{center}}
\newcommand{\ec}{\end{center}}
\newcommand{\bi}{\begin{itemize}}
\newcommand{\ei}{\end{itemize}}
\newcommand{\ben}{\begin{enumerate}}
\newcommand{\een}{\end{enumerate}}
\newcommand{\bd}{\begin{definition}}
\newcommand{\ed}{\end{definition}}
\def\beq{\vskip .1cm \begin{equation}}
\def\eeq{\end{equation} \vskip .1cm \noindent}
\def\beqn{\vskip .1cm \begin{eqnarray}}
\def\eeqn{\end{eqnarray} \vskip .1cm \noindent}
\def\beqnn{\vskip .01cm \begin{eqnarray*}}
\def\eeqnn{\end{eqnarray*} \vskip .01cm \noindent}
\def\bcase{\vskip .1cm \begin{numcases}}
\def\ecase{\end{numcases} \vskip .1cm \noindent}
\def\bsbcase{\vskip .1cm \begin{subnumcases}}
\def\esbcase{\end{subnumcases} \vskip .1cm \noindent}

\def\defeq{{:=}}%{{\stackrel{\Delta}{=}}}
\newtheorem{theorem}{Theorem}
\newtheorem{corollary}{Corollary}
\newtheorem{lemma}{Lemma}
\newtheorem{definition}{Definition}

\newenvironment{myproof}{\noindent{\em Proof:} \hspace*{1em}}{
    \hspace*{\fill} $\Box$ }
\newenvironment{proof_of}[1]{\noindent {\em Proof of #1:}
    \hspace*{1em} }{\hspace*{\fill} $\Box$ }

\def\psfancypar#1#2{\begingroup\def\par{\endgraf\endgroup\lineskiplimit=0pt}
               \setbox2=\hbox{\large\sc #2}
               \newdimen\tmpht \tmpht \ht2 \advance\tmpht by \baselineskip
               \font\hhuge=Times-Bold at \tmpht
               \setbox1=\hbox{{\hhuge #1}}
               \count7=\tmpht \count8=\ht1
               \divide\count8 by 1000 \divide\count7 by \count8 
               \tmpht=.001\tmpht\multiply\tmpht by \count7 
               \font\hhuge=Times-Bold at \tmpht
               \setbox1=\hbox{{\hhuge #1}}
               \noindent
                \hangindent1.05\wd1
               \hangafter=-2 {\hskip-\hangindent
               \lower1\ht1\hbox{\raise1.0\ht2\copy1}%
                \kern-0\wd1}\copy2\lineskiplimit=-1000pt}

\def\Kout{\setbox1=\hbox{\Huge\bf K}\hbox to
1.05\wd1{\hspace{.05\wd1}
}}
\def\Sout{\setbox1=\hbox{\Huge\bf S}\hbox to 1.05\wd1{\hspace{.05\wd1}
}}

\def\Etot{\mbox{C}}

\def\predecessor{Pred}

\def\E{\mathcal{E}}
\def\nbd{\mathcal{N}}

\def\mst{\mbox{MST}}
\def\nng{\mbox{NNG}}
\def\lcg{\mbox{DTG}}

\def \spt{{\mbox{SP}}}

\def\dfmrf{{\mbox{DFMRF}}}

\def\snng{{\mbox{\scriptsize NNG}}}
\def\smst{{\mbox{\scriptsize MST}}}

\def\dmst{{\mbox{DMST}}}

\def\lcg{\mbox{FG}}

\def \spt{{\bar{C}}}
\def\predecessor{\Nc_p}%{ImmPred}

\def\nbd{\Nc_u}%{Nbd}

\def\pot{\phi}

\def\Proc{Proc}

\def\E{E}

\def\dfnng{\mbox{AggMST}}

\newcommand{\mbbE}{\mathbb{E}}
\newcommand{\Bmsc}{Q}
\newcommand{\Vmsc}{\bfV}%{\mathscr{V}}
\newcommand{\Lmsc}{\mathscr{L}}

\def\approxratio{\rho}

\def\constgd{{\zeta}}
\def\dep{{\Gmsc}}
\def\sdep{{\Gmsc}}

\def\nbd{{\mathcal{N}}}

\def\Lmsc{{E}}
\def\predecessor{{\mathcal{N}_p}}

\def\Cc{{\Cmsc}}
\def\Bc{{\Bmsc}}
\def\Proc{{\mbox{Proc}}}
\def\Etot{{\Emsc}}

\def\Bc{{Q}}
\def\spt{{\Emsc^{\mbox{\tiny SP}}}}

\def\comm{{\mathscr{N}_g}}
\def\complete{C_g}

\def\N{\mathbb{N}}
\def\E{\mathbb{E}}

\def\0{{\bf 0}}

\def\R{\mathbb{R}}

\renewcommand{\E}{\mathbb E \,}

\def\la{{\lambda}}

\renewcommand{\P}{{{\cal P}}}

\newcommand{\A}{{\cal A}}

\def\bdm{\begin{displaymath}}
\newcommand{\edm}{\end{displaymath}}
\def\benu{\begin{enumerate}}
\def\eenu{\end{enumerate}}
\def\be{\begin{equation}}
\def\ee{\end{equation}}
\def\bea{\begin{eqnarray}}
\def\eea{\end{eqnarray}}
\newcommand{\bean}{\begin{eqnarray*}}
\newcommand{\eean}{\end{eqnarray*}}
\newcommand{\bear}{\begin{eqnarray}}
\newcommand{\eear}{\end{eqnarray}}

\def\R{\mathbb{R}}

\def\A{{\cal A}}

\def\la{{\lambda}}

\def\pl{\parallel}

\title{Energy Scaling Laws for Distributed Inference\\ in Random Fusion  Networks}
\author{Animashree Anandkumar$^\dagger$\thanks{$^\dagger$Corresponding
author.}, ˜\IEEEmembership{Student~Member,~IEEE}, Joseph E. 
Yukich,\\Lang Tong, ˜\IEEEmembership{Fellow,~IEEE}, and Ananthram 
Swami,~\IEEEmembership{Fellow,~IEEE}
\thanks{\scriptsize
A. Anandkumar and L. Tong  are with the School of Electrical and
Computer Engineering, Cornell University, Ithaca, NY 14853, USA.
 Email: {\tt\{aa332@,ltong@ece.\}cornell.edu}}
 \thanks{\scriptsize J.E. Yukich is with the Department of Mathematics,
 Lehigh University, Bethlehem, Pa. 18015.
E-mail:  {\tt joseph.yukich@lehigh.edu}.}
 \thanks{\scriptsize A. Swami is with the Army Research Laboratory, Adelphi, MD 20783 USA
E-mail: {\tt a.swami@ieee.org}.}
\thanks{\scriptsize This work was
supported in part through collaborative participation in 
Communications and Networks Consortium sponsored by the U.~S. Army 
Research Laboratory under the Collaborative Technology Alliance 
Program, Cooperative Agreement DAAD19-01-2-0011 and  by the Army 
Research Office under Grant ARO-W911NF-06-1-0346. The first author 
is supported by the IBM Ph.D Fellowship for the year 2008-09 and is 
currently a visiting student at  MIT, Cambridge, MA 02139. The 
second author was partially supported  by NSA grant H98230-06-1-0052 
and NSF grant DMS-0805570. The U. S. Government is authorized to 
reproduce and distribute reprints for Government purposes 
notwithstanding any copyright notation thereon.}\thanks{\scriptsize 
Manuscript received on 08/25/2008, revised  on 02/01/2009. Parts of 
this paper were presented at 
\cite{Anandkumar&etal:08JSM,Anandkumar&etal:08Allerton}}}
\begin{document}

\maketitle
\begin{abstract}
The energy scaling laws of multihop data fusion networks for 
distributed inference are considered.  The fusion network   consists 
of randomly located sensors distributed i.i.d. according to a 
general spatial distribution in an expanding region.  Under  Markov 
random field (MRF)  hypotheses, among the class of data-fusion 
policies which enable optimal statistical inference at the fusion 
center using all the sensor measurements, the policy with the  
minimum average energy consumption is bounded below by the average 
energy of fusion along the minimum spanning tree,  and above by a 
suboptimal policy, referred to as Data Fusion for Markov Random 
Fields (DFMRF). Scaling laws are derived for the energy consumption 
of the optimal and suboptimal fusion policies.  It is shown that  
the average asymptotic energy of the DFMRF scheme is strictly  
finite  for a class of MRF models with Euclidean {\em stabilizing} 
dependency graphs. 
\end{abstract}

\begin{IEEEkeywords} Distributed inference,  graphical models,
  Euclidean random  graphs, stochastic geometry and data fusion.\end{IEEEkeywords}

\vspace{1em}

\section{Introduction}\label{sec:intro}

\IEEEPARstart{W}e consider the problem of distributed statistical 
inference in a network of randomly located sensors  taking  
measurements  and transporting the locally processed data to a  
designated fusion center. The fusion center then makes an inference 
about the underlying phenomenon based on the data collected from all 
the sensors.

For statistical inference using wireless sensor networks, energy 
consumption is an important design parameter.  The transmission 
power required to reach  a receiver   distance $d$ away  with  a 
certain signal-to-noise ratio (SNR)     scales in the order of 
$d^\nu$, where $2\le \nu \le 6$ is the path loss 
\cite{Ephremides:02Wcom}. Therefore, the cost of moving data from 
sensor locations to the fusion center, either through direct 
transmissions or through multihop forwarding, significantly affects 
the lifetime of the network.

\subsection{Scalable data fusion}\label{sec:intro_scalable}

We investigate the cost of data fusion for inference, and its 
scaling behavior with the size of the network and the area of 
deployment. In particular, for a network of $n$ random sensors 
located at points $\Vmsc_n=\{V_1,\cdots, V_n\}$  in $\R^2$, a {\em 
fusion policy} $\pi_n$ maps $\Vmsc_n$ to a set of scheduled 
transmissions and computations. The average cost (e.g., energy)  of 
a policy is given by \beq\bar{\Ec}(\pi_n(\Vmsc_n))\defeq \frac{1}{n} 
\sum_{i\in \Vmsc_n} \Ec_i(\pi_n(\Vmsc_n)),\label{eq:En} \eeq where 
$\Ec_i(\pi_n(\Vmsc_n))$ is the cost at node $i$ under policy 
$\pi_n$. The above average cost  is random, and we are interested in 
its  scalability in random networks as $n\rightarrow \infty$.

\bd[Scalable Policy]  A sequence of policies 
$\pi\defeq(\pi_n)_{n\geq 1}$ is scalable on
  average if
\[\lim_{n \to \infty} \mbbE(\bar{\Ec}(\pi_n(\Vmsc_n))) =
\bar{\Ec}_\infty(\pi)< \infty\] where the expectation $\mbbE$ is 
with respect to the random locations $\Vmsc_n$, and 
$\bar{\Ec}_\infty(\pi)$ is referred to as the {\em scaling 
constant}.
 A sequence of policies $\pi_n$ is {\em weakly scalable}
if
\[\plim_{n\to\infty} \bar{\Ec}(\pi(\Vmsc_n))) = \bar{\Ec}_\infty(\pi)
<\infty,\]where $\plim$ denotes convergence in probability. It is 
{\em strongly scalable} if the above average energy converges almost 
surely and is {\em $L^2$ (mean-squared) scalable} if the convergence 
is in mean square. \ed

Hence, a scalable fusion policy implies a  finite average   energy 
expenditure even as the network size increases. We  focus mostly on 
the $L^2$ scalability of the fusion policies, which implies  weak  
and average scalability \cite{Billingsley:book}. Further, we are 
interested in {\em lossless} data-fusion policies which enable  the 
fusion center   to perform optimal statistical inference with the 
best inference accuracy {\em as if} all the raw sensor data   were 
available.

To motivate this study, first consider two simple fusion policies: 
the direct transmission policy (DT) in which all sensors transmit 
directly to the fusion center  (single hop), and the shortest-path 
(SP)  policy, where each node forwards its raw data to the fusion 
center using the shortest-path route without any data combination at 
the intermediate nodes.

We assume, for now, that $n$ sensor nodes are uniformly distributed
  in a square of area
$ n $. It is perhaps not surprising that neither of the above two 
policies is scalable as  $n\rightarrow \infty$.  For the DT 
policy\footnote{The direct transmission policy may not even be 
feasible, depending on the maximum transmission power constraints at 
the sensors.}, intuitively, the average transmission range from the 
sensors to the fusion center  scales as $\sqrt{n}$,  thus 
$\bar{\Ec}(\mbox{DT}(\Vmsc_n))$ scales as $n^\frac{\nu}{2}$. On the 
other hand, we expect the SP policy to have better scaling since it 
chooses the best multi-hop path to forward data from each node to 
the fusion center. However, even in this case,  there is no finite 
scaling. Here, the average number of hops in the shortest path from 
a node to the fusion center scales in the order of $\sqrt{n}$, and 
thus, $\bar{\Ec}(\mbox{SP}(\Vmsc_n))$ scales in the order of 
$\sqrt{n}$. Rigorously establishing the scaling laws for these two 
non-scalable policies is not crucial at this point since the same 
scaling laws can be easily established for regular networks when 
sensor nodes are on  two-dimensional lattice points. See 
\cite{Li&Dai:08SPL}.

Are there scalable policies for data fusion?  Among all the fusion 
policies not performing data combination at the intermediate nodes,   
the shortest-path (SP) policy minimizes the total energy. Thus, no 
scalable policy exists unless nodes cooperatively combine their 
information, a process known as {\em data  aggregation}.  Data 
aggregation, however, must be considered in conjunction with the 
performance requirements of specific applications. In this paper, we 
assume that  optimal statistical inference is performed at the 
fusion center {\em as if }   all the raw sensor data were available, 
and this places a constraint on data aggregation. For instance, it 
rules out sub-sampling of the sensor field, considered in   
\cite{Anandkumar&etal:09INFOCOM}.

\subsection{Summary of results and contributions}
In this paper, we investigate the energy scaling laws of lossless 
fusion policies which  are allowed to perform     data aggregation 
at the intermediate nodes, but ensure that the fusion center 
achieves the same  inference accuracy {\em as if} all the raw 
observations were collected without any data combination. We assume 
that the underlying binary hypotheses for the sensor measurements 
can be modeled as Markov random fields  (MRF).

For   sensor locations $\Vmsc_n$ and possibly correlated sensor 
measurements, finding the minimum energy fusion policy under the 
constraint of optimal inference is given by 
\begin{equation} \Etot(\pi^*(\bfV_n))=\inf_{\pi \in 
\mathfrak{A}} \sum_{i\in \bfV_n} \Ec_i(\pi(\bfV_n)) 
\label{eq:optimal},
\end{equation}
where $\mathfrak{A}$ is the set of valid lossless data-fusion 
policies
\[
\mathfrak{A} \defeq \{\pi : \mbox{optimal inference is achieved   at 
the fusion center}\}.
\] In general, the above optimization is NP-hard 
\cite{Anandkumar&etal:08INFOCOM}, and  hence, studying its  energy 
scaling behavior directly is intractable.  We establish upper and 
lower bounds on the   energy of this optimal policy  $\pi^*$ and 
analyze the  scaling behavior  of these bounds. The lower bound is 
obtained via  a policy conducting fusion along the Euclidean minimum 
spanning tree (MST), which is shown to be optimal when the sensor 
measurements are statistically independent under both hypotheses. 
The upper bound on the optimal fusion policy is established through 
a specific suboptimal fusion policy, referred to as Data Fusion over 
Markov Random Fields (DFMRF). DFMRF becomes optimal when 
observations are  independent under either hypothesis, where it 
reduces to fusion along the MST. For certain spatial dependencies 
among sensor measurements of practical significance, such as the 
Euclidean 1-nearest neighbor graph,  DFMRF has an approximation 
ratio $2$, \ie its energy is no more than twice that of the optimal 
fusion policy, independent of the size and configuration of the 
network.

We then proceed to establish a number of asymptotic properties of 
the $\dfmrf$ policy in Section~\ref{sec:scaling}, including its 
energy scalability, its performance bounds, and the approximation 
ratio with respect to the optimal fusion policy  when the sensor
 measurements have dependencies described by a  $k$-nearest neighbor
graph or a disc  graph (continuum percolation). Applying techniques 
developed in \cite{Steele:88,Yukich:00,Aldous&Steele:04, 
Penrose&Yukich:03AAP}, we provide a precise characterization of the 
scaling bounds as a function of sensor  density and sensor placement 
distribution.  These asymptotic bounds for DFMRF, in turn, imply 
that the  optimal fusion policy is also scalable. Hence, we use the 
DFMRF policy as  a vehicle to establish scaling laws for optimal 
fusion. Additionally, we use the energy scaling constants to 
optimize the distribution of the sensor placements. For   
independent measurements conditioned on each hypothesis, we show 
that the uniform distribution of the sensor nodes minimizes the 
asymptotic average energy consumption over all i.i.d spatial 
placements when the path-loss exponent of transmission is greater 
than two   $(\nu > 2)$.  For $\nu \in [0, 2)$, we show that  the 
uniform distribution is, in fact, the most expensive\footnote{The 
path-loss exponent for wireless transmissions  satisfies $\nu > 2$.} 
node configuration in terms of routing costs.   We further show that 
the optimality of the uniform node distribution applies for both the 
lower and  upper   bounds on the average energy consumption of the 
optimal fusion policy under Markov random field measurements with 
$k$-nearest neighbor dependency graph or the disc dependency graph 
under certain conditions.

To the best of our knowledge, our results are the first to establish 
the energy scalability of data fusion for certain  correlation 
structures of the sensor measurements.  The use of   energy scaling 
laws for the design of efficient sensor placement is new and  has 
direct engineering implications. The fusion policy DFMRF first 
appeared in \cite{Anandkumar&Tong&Swami:07CISS}, and is made precise 
here with detailed asymptotic analysis using the weak law of large 
numbers (WLLN) for {\em stabilizing} Euclidean graph functionals. 
One should  not expect that scalable data fusion is always possible, 
and  at the end of Section \ref{sec:scaling}, we  discuss examples 
of correlation structures where scalable lossless data-fusion policy 
does not exist.

\subsection{Prior and related work}

%no concept of rate here since one shot transmission. Also scheduling
% is not an issue

The seminal work of Gupta and Kumar \cite{Gupta&Kumar:00IT} on the 
capacity of wireless networks has stimulated extensive studies 
covering a broad range of networking problems with different 
performance metrics. See also \cite{Franceschetti&Meester:book}. 
Here, we  limit ourselves  to the  related works  on energy 
consumption and data fusion for statistical inference.

Results on scaling laws for energy consumption are limited. In 
\cite{Zhao&Tong:05JSAC}, energy scaling laws for multihop wireless 
networks (without any data fusion) are derived under different 
routing strategies. The issue of node placement for desirable energy 
scaling has been considered in 
\cite{Liu&Haenggi:06TranPDS,WuChenDas:TranPDS}, where it is argued 
that  uniform node placement, routinely considered in the 
literature, has poor energy performance when there is no data 
fusion.  It is interesting to note that, for  fusion networks,   
uniform sensor distribution is in fact optimal among a general class 
of distributions.  See Section~\ref{sec:dfmrf_scaling}.

Energy-efficient data fusion has received a great deal of attention 
over the past decade. See a few recent surveys in 
\cite{ZhaoSwamiTong:06SPM,Giridhar&Kumar:06CommMag}. It has  been 
recognized that sensor observations tend to be correlated, and that 
correlations should be exploited through data fusion. One line of 
approach is the use of distributed compression with the aim of 
reconstructing all the measurements at the fusion center. Examples 
of such approaches can be found in 
\cite{Cristescu&etal:06TON,Rickenbach&Wattenhofer:04FMC,gupta2005egc}.

While sending data from all sensors to the fusion center is 
certainly sufficient to ensure optimal inference, it is not 
necessary. More relevant to our work is the idea of data 
aggregation, \eg 
\cite{madden2005taq,Intanagonwiwat&etal:00MOBICOM,Krishnamachari&Estrin&Wicker:02INFOCOM}. 
Finding aggregation policies for correlated data, however, is 
nontrivial; it depends on the specific applications for which the 
sensor network is designed.  Perhaps a more precise notion of 
aggregation is in-network function computation  where  certain 
functions are computed by passing intermediate values among  nodes 
\cite{Giridhar&Kumar:05IPSN,Giridhar&Kumar:05JSAC,Subramaniam&Gupta&Shakkotai:07ISIT,Ayaso&Shah&Dahleh:08ISIT}. 
However, these works are mostly concerned with computing symmetric 
functions such as the sum function, which in general, do not satisfy 
the constraint of optimal statistical inference at the fusion 
center.

%, and with the rate of function computation, while we are concerned 
%with one-time computation for inference.

In the context of statistical inference using wireless sensor
networks, the idea of aggregation and in-network processing has been
explored by several authors. See
\cite{Sung&Misra&Tong&Ephremides:07JSAC,%
Chamberland&Veeravalli:06IT,Misra&Tong:08SP,Sung&etal:08SP,Sung&etal:08ISIT,%QuanKaiserSayed:07IPSN,Yang&Blum:07IPSN
Katenka&Levina&Michailidis:JRCSQIT,Yu&Ephremides:06IPSN}. Most
relevant to
our work are \cite{Sung&Misra&Tong&Ephremides:07JSAC,Sung&etal:08ISIT,%
Chamberland&Veeravalli:06IT,Misra&Tong:08SP,Sung&etal:08SP} where 
the Markovian correlation structures  of sensor measurements are 
exploited explicitly. These results mostly deal with one-dimensional 
node placements, and  do not deal with randomly placed nodes or 
energy scaling laws.

%The use of the MRF model for spatial data in sensor networks is
%relatively new (\eg \cite{Willsky:bookchapter}), although it is
%widely used  in  geo-statistics  \cite{Cressie:book,pettitt2002cag,
%Rue&Held:book}. This could be due to the complexity of the model for
%arbitrarily-placed nodes.

The results presented in this paper extend some of our earlier work 
in the direction of scaling-law analysis in random fusion networks. 
In 
\cite{Anandkumar&Tong&Swami:07CISS,Anandkumar&etal:08INFOCOM,Anandkumar&etal:bookchapter}, 
for fixed network size and  node placement, we analyzed the minimum 
energy fusion policy for optimal inference and showed that it 
reduces to the {\em Steiner-tree} optimization problem under certain 
constraints. We also proposed a suboptimal fusion policy called  the 
DFMRF\footnote{The DFMRF policy is referred to as $\dfnng$ in 
\cite{Anandkumar&etal:08INFOCOM,Anandkumar&etal:bookchapter}.}. In 
\cite{Anandkumar&Tong&Swami:08SP}, we analyzed the optimal sensor 
density for uniform node placement which maximizes the inference 
error exponent under an average energy constraint,  and in 
\cite{Anandkumar&Tong&Swami:09IT,Anandkumar&etal:09ISIT}, we derived   
the error exponent for MRF hypotheses. In 
\cite{Anandkumar&etal:09INFOCOM}, we analyzed optimal sensor 
selection (\ie sub-sampling) policies for achieving tradeoff between 
fusion costs and inference performance.

The energy scaling laws derived in this paper rely heavily on 
several results on the law of large numbers for geometric random 
graphs.  We have extensively  borrowed the formulations and 
techniques of Penrose and Yukich 
\cite{Penrose&Yukich:02AAP,Penrose&Yukich:03AAP}.  See Appendix 
\ref{sec:lln} for a brief description and 
\cite{Steele:88,Yukich:00,Penrose:book} for detailed expositions of 
these ideas.

%%%%%%%%%%%%%%%%
% Model and assumptions
%%%%%%%%%%%%%%%%%%%%
\section{System Model}

In this paper, we  consider various graphs. Chief among these are 
(i)  {\em dependency} graphs specifying the correlation structure of 
sensor measurements, (ii) {\em network} graphs denoting the 
(directed) set of feasible  links for communication, and (iii)  {\em 
fusion} policy digraphs denoting the (directed) links used by a 
policy to route and aggregate data according to a given sequence. 
Note that the fusion policy takes the dependency graph and the 
network graph as inputs and outputs the fusion-policy digraph. The 
dependency and network graphs can be independently specified and are 
in general, functions of the  sensor locations.

%%%%%%%%%%%%%%%%
% Model and assumptions
%%%%%%%%%%%%%%%%%%%%
\subsection{Stochastic model of sensor locations}\label{sec:location}
We assume that $n$ sensor nodes (including the fusion center) are
placed randomly with sensor $i$ located at $V_i \in \R^2$. By
convention, the  fusion center is  denoted by  $i=1$, and is located 
at $V_1\in \R^2$. We denote the set of locations of the $n$ sensors  
by $\Vmsc_n\defeq\{V_1,\ldots,V_n\}$. For our scaling law analysis, 
we consider a sequence of sensor populations placed in  expanding 
square regions $\Bmsc_{\frac{n}{\lambda}}$ of area 
$\frac{n}{\lambda}$ and centered at the origin $\0\in\R^2$, where we 
fix $\lambda$ as the overall sensor density and let the number of 
sensors $n\rightarrow \infty$.

To generate sensor locations $V_i$, first let $\Bmsc_1:=
[-\frac{1}{2},\frac{1}{2}]^2$ be the unit-area square\footnote{The 
results in this paper hold for $\tau$ defined on any convex unit 
area.}, and $X_i \overset{i.i.d.}{\sim} \tau, 1\leq i \leq n,$ be a 
set of $n$ independent and identically distributed (i.i.d.) random 
variables distributed on support $\Bmsc_1$ according to $\tau$. 
Here, $\tau$ is a probability density function (pdf) on $\Bmsc_1$ 
which is bounded away from zero and infinity.  We then generate 
$V_i$ by scaling $X_i$ accordingly: $V_i=\sqrt{\frac{n}{\la}}X_i \in 
\Bmsc_{\frac{n}{\lambda}}$. A useful special case is the  uniform 
distribution $(\tau\equiv 1)$. Let $\Pc_{a}$ be the homogeneous 
Poisson distribution   on $\R^2$ with intensity $a>0$.
%
%Technically, we have a marked point process for the sensors,  where
%the marks are values of the sensor measurements. However, in this
%paper, we only evaluate  functionals incorporating the {\em
%dependency graph} of the measurements. This  is a property of the
%joint pdf of the measurements, and not the values of the
%measurements themselves.

\subsection{Graphical inference model: dependency
graphs}\label{sec:dependency}  We consider  the statistical 
inference problem of simple binary hypothesis testing, $\Hc_0$ vs. 
$\Hc_1$, on a pair of Markov random fields.    Under regularity 
conditions \cite{Bremaud:book}, a MRF is defined by its (undirected) 
dependency graph $\Gmsc$ and an associated pdf $f(\cdot\mid\Gmsc)$.

Under hypothesis $\Hc_k$ and sensor location set 
$\Vmsc_n=\{V_1,\cdots, V_n\}$ generated according to the stochastic 
model in Section~\ref{sec:location}, we assume that the dependency 
graph $\Gmsc_k:=(\Vmsc_n,\Lmsc_k)$   models the correlation among 
the sensor observations.  Note that
 the node location set $\bfV_n$ under the two hypotheses are identical.
Set $\Lmsc_k$ is the set of   edges of the dependency graph 
$\Gmsc_k$,  and it defines the correlations of the sensor 
observations, as described in the next section. 

%One can also describe the dependency graph $\Gmsc_k$ through
%$\Vmsc$ and the collection $\Cmsc_k$ of maximal cliques
% specified by links in $\Lmsc_k$.

We restrict our attention to {\em proximity}-based Euclidean 
dependency graphs. In particular, we consider  two classes of 
dependency graphs\footnote{The $k$-nearest neighbor graph 
$(k$-$\nng)$ has edges $(i,j)$ if $i$ is one of the top   $k$ 
nearest neighbors of $j$ or viceversa, and ties are arbitrarily 
broken. The disc graph has edges between any two points within a 
certain specified Euclidean distance (radius).}:  the (undirected) 
$k$-nearest neighbor graph ($k$-NNG) and the  disc graph, also known 
as the  continuum percolation graph. We expect that our results 
extend to other locally-defined dependency structures such as  the 
Delaunay, Voronoi, the minimum spanning tree, the sphere of 
influence  and the Gabriel graphs. An important 
  property of the aforementioned graphs is a certain 
{\em stabilization} property (discussed in Appendix \ref{sec:lln}) 
facilitating asymptotic scaling analysis.

%We can justify the choice of above proximity-based graphs for
%dependency since typically,  many spatial phenomena such as rainfall
%and temperature data  are based on proximity, and are well-modeled
%with the $k$-NNG or the disc graph, depending on the scenario
%\cite{pettitt2002cag,Cressie:book}.

\subsection{Graphical inference model: likelihood
functions}\label{sec:mrf}  We denote the   measurements from all the 
$n$ sensors placed at   fixed locations $\bfv_n$ by $\bfY_{\bfv_n}$. 
The statistical inference problem can now be stated as the following 
hypothesis test:\begin{align} \Hc_0:& [\bfY_{\Vmsc_n}, \Vmsc_n] \sim 
f(\bfy_{\bfv_n} \mid\Gmsc_0(\bfv_n), \Hc_0)\prod_{i=1}^n 
\tau(\sqrt{\tfrac{\lambda}{n}}v_i),\nn\\\Hc_1:& [\bfY_{\Vmsc_n}, 
\Vmsc_n] \sim f(\bfy_{\bfv_n} \mid\Gmsc_1(\bfv_n), 
\Hc_1)\prod_{i=1}^n 
\tau(\sqrt{\tfrac{\lambda}{n}}v_i),\label{eqn:hyp}\end{align} where 
$f(\bfy_{\bfv_n} \mid\Gmsc_k, \Hc_k)$ is the pdf of $\bfY_{\bfv_n}$ 
given the  dependency graph $\Gmsc_k(\bfv_n)$
 under hypothesis $\Hc_k$.  Note that the sensor locations $\bfV_n$ have the same distribution under
  either hypothesis.  Therefore, only the conditional
  distribution of $\bfY_{\bfv_n}$ given the sensor locations  $\bfV_n=\bfv_n$ under each hypothesis is
relevant for inference.

Under each hypothesis, the dependency graph specifies 
conditional-independence relations between the sensor  measurements 
\cite{Bremaud:book} \begin{equation}Y_i \indep 
\bfY_{\Vmsc_n\setminus \nbd(i;\Gmsc_k)}\mid 
\{\bfY_{\nbd(i;\Gmsc_k)},\Vmsc_n\},\quad \mbox{under 
}\Hc_k,\end{equation} where $\nbd(i;\Gmsc_k)$ is the set of 
neighbors of $i$ in $\Gmsc_k$, and $\indep$ denotes conditional 
independence. In words,  the measurement at a node is conditionally 
independent of the rest of the network, given the node locations 
$\bfV_n$ and the  measurements at its neighbors   in the dependency 
graph.

The celebrated  Hammersley-Clifford theorem \cite{Clifford:1990mrf} 
states that, under the positivity condition\footnote{The positivity 
condition rules out degeneracy among a subset of nodes: 
$Y_1=Y_2\ldots=Y_k$, where it is not required for every node to 
transmit at least once for computation of likelihood ratio.},
 the log-likelihood function of a MRF with  dependency graph $\Gmsc_k$ can be expressed as
\begin{equation}
-\log f(\bfy_{\bfv_n}\mid\Gmsc_k(\bfv_n), \Hc_k) = \sum_{c \in 
\Cmsc_k} \psi_{k,c}(\bfy_c),\quad k=0,1, \label{eqn:hammersley}
\end{equation}
where $\Cmsc_k$ is a collection of  (maximal) cliques\footnote{A 
clique is a complete subgraph, and a maximal clique is a clique 
which is not contained in a bigger clique.} in $\Gmsc_k(\bfv_n)$, 
the functions $\psi_{k,c}$, known as   {\em clique potentials}, are 
real valued,  and not zero everywhere on the support of the 
distribution of $\bfy_c$. 

We assume that the normalization constant (partition function) is 
already incorporated in the potential functions to ensure that 
(\ref{eqn:hammersley}) indeed describes a probability measure. In 
general, it is NP-hard to evaluate the normalization constant given 
arbitrary potential functions \cite{Jerrum&Sinclair:93}, but  can be 
carried out at the fusion center without any need for communication 
of sensor measurements.

% not clear if uniqueness needs to be mentioned.

%The  case of   independent observations is of special
%interest.  In this case, the corresponding dependency graph $\Gmsc$
%contains no edge, and the set of maximal cliques is $\Vmsc$ itself.
%Thus the log-likelihood function is given by
%\[ -\log f(\bfY_\Vmsc\mid\Gmsc_k,\Hc_k) =- \sum \log f_i(Y_i\mid\Hc_k)
%\]
%where $f_i$ is the marginal pdf for $Y_i$.

\subsection{Communication model  and energy
consumption}\label{sec:energy}

The set of feasible communication links form the (directed) {\em 
network graph} denoted by $\comm(\bfv_n)$, for a given realization 
of sensor locations $\bfV_n = \bfv_n$. We assume that it is 
connected  but not necessarily fully connected, and that it contains 
the Euclidean minimum spanning tree over the node set $\bfv_n$ and 
directed towards the fusion center $v_1$, denoted by 
$\dmst(\bfv_n;v_1)$.  Usually in the literature,  in order to 
incorporate the maximum power constraints at the nodes, the network 
graph is assumed to be a disc graph with radius above the 
connectivity threshold \cite{Franceschetti&Meester:book}, but we do 
not limit to this model. Transmissions are scheduled so as to not 
interfere with one other. Nodes are capable of adjusting their 
transmission power depending on the location of the receiver.

A fusion policy $\pi(\bfv_n)$ consists of a  transmission schedule 
with the transmitter-receiver pairs and the aggregation algorithm 
that allows a node to combine its own and received values  to 
produce a new communicating value. We model a fusion policy $\pi$  
by a   {\em fusion-policy digraph}, $\Fmsc_\pi:= 
(\bfv_n,\overrightarrow{\Lmsc}_\pi)$,  and 
$\overrightarrow{\Lmsc}_{\pi}$ contains {\em directed links}.
 A directed\footnote{We denote a directed link by $\langle i,j\rangle$ and an
 undirected link by $(i,j)$.} link $\langle i,j\rangle$ denotes a direct transmission from $i$ to $j$ and
 is required to be a member in the network graph $\comm(\bfv_n)$ 
 for   transmissions to be feasible.  If
one node communicates with another node $k$ times, $k$ direct links 
are present between these two nodes in the edge set 
$\overrightarrow{\Lmsc}_\pi$ of the fusion policy $\pi$. Since we 
are only interested in characterizing the overall energy 
expenditure, the order of transmissions  is not important; we only 
need to consider the associated cost with each  link in 
$\overrightarrow{\Lmsc}_{\pi}$
 and calculate the sum cost for $\pi$.

Nodes communicate in the form of packets. Each packet contains bits
for at most one (quantized) real variable and other overhead bits
independent of the network size.  We assume that all real 
variables\footnote{In principle, the raw and aggregated data may 
require different amount of energy for communication, and   can be  
incorporated into our framework.} are quantized to $K$ bits, and $K$ 
is independent of network size and is sufficiently large that 
quantization errors can be ignored. Thus, for node $i$ to transmit 
data to node $j$ which is distance $|i,j|$ away, we assume that node 
$i$   spends energy\footnote{Since nodes only communicate a finite 
number of bits,
 we use energy instead of power as the cost measure.}  $\gamma |i,j|^\nu$.
 Without loss of generality, we
assume $\gamma=1$. Hence, given a fusion policy
$\Fmsc_\pi=(\bfv_n,\overrightarrow{\Lmsc}_\pi)$ of network size $n$, 
the average energy consumption is  given by
\begin{equation}
\bar{\Ec}(\pi(\bfv_n))=\frac{1}{n} \Ec(\pi(\bfv_n))=\frac{1}{n} 
\sum_{\langle i,j\rangle \in \overrightarrow{\Lmsc}_\pi}  |i,j|^\nu, 
\quad 2\leq \nu\leq 6.\label{eqn:fusiondigraph}
\end{equation}

The model specification is now complete.

%%%%%%%
% Optimal
%%%%%%%

\section{Minimum Energy Data Fusion}\label{sec:opt}
 In this section, we present  data-fusion policies aimed at minimizing energy 
expenditure under the constraint of optimal statistical inference at 
the fusion center, given in \eqref{eq:optimal}.   The scalability of 
these  policies is deferred to Section~\ref{sec:scaling}.

\subsection{Optimal data fusion: a reformulation}\label{sec:fusion_reformulation}

We   consider  correlated sensor measurements under the Markov 
random field model.  The inference problem, defined in 
(\ref{eqn:hyp}), involves two different graphical models, each with 
its own dependency graph and associated likelihood function.  They 
do share the same node location set $\Vmsc_n$ which allows us to 
merge the two graphical models into one.

For a given realization of sensor locations $\bfV_n = \bfv_n$, 
define the {\em joint} dependency graph $\Gmsc\defeq(\bfv_n,E),$ 
where $ E\defeq E_0\bigcup E_1$, as the union of the two (random) 
dependency graphs $\Gmsc_0$ and $\Gmsc_1$. The minimal sufficient 
statistic\footnote{A sufficient statistic is a well-behaved function 
of the data, which is as informative as the raw data for inference. 
It is minimal if it is a function of every other sufficient 
statistic \cite{Poor:book}.} is given by the log-likelihood ratio 
(LLR)  \cite{Dynkin:61}. With the substitution of 
(\ref{eqn:hammersley}), it is given by
\begin{eqnarray}
L_\Gmsc(\bfy_{\bfv_n}) &\defeq& \log \frac{f(\bfy_{\bfv_n}
\mid\Gmsc_0(\bfv_n),\Hc_0)}{f(\bfy_{\bfv_n}\mid\Gmsc_1(\bfv_n),\Hc_1)}\nn\\
&=& \sum_{a \in \Cmsc_{1}}\psi_{1,a}(\bfy_{a})-\sum_{b
\in \Cmsc_{0}}\psi_{0,b}(\bfy_{b}) \nn\\
&\defeq& \sum_{c \in
\Cmsc}\phi_{c}(\bfy_{c}),~~~\Cmsc\defeq\Cmsc_0\bigcup
\Cmsc_1,\label{eqn:LLR}
\end{eqnarray}where $\Cmsc$ is the set of maximal cliques in $\Gmsc$ and the 
effective potential functions $\pot_c$ are given by \beq 
\pot_c(\bfy_c) \defeq\!\!\!\sum_{a \in \Cmsc_{1},a\subset 
c}\!\!\!\psi_{1,a}(\bfy_{a})- \!\!\!\sum_{b \in \Cmsc_{0},b \subset 
c}\!\!\!\psi_{0,b}(\bfy_{b}),\quad \forall\, c \in \Cmsc.\eeq 
Hereafter, we  work with $(\Gmsc,L_\Gmsc(\bfy_{\bfv_n}))$ and refer 
to the joint dependency graph $\Gmsc$ as just the dependency graph.  

Note that the log-likelihood ratio is minimally sufficient  
\cite{Dynkin:61} (\ie maximum dimensionality reduction) implying 
maximum possible savings in routing energy through aggregation under 
the constraint of optimal statistical inference. Given a fixed 
node-location set $\bfv_n$, we can now reformulate the optimal 
data-fusion problem in \eqref{eq:optimal} as the following 
optimization
\begin{equation}
\Etot(\pi^*(\bfv_n))=\inf_{\pi \in \mathfrak{F}_\Gmsc} \sum_{i\in 
\bfv_n} \Ec_i(\pi(\bfv_n)) \label{eq:optimal_reform},
\end{equation}
where $\mathfrak{F}_\Gmsc$ is the set of valid data-fusion policies
\[
\mathfrak{F}_\Gmsc \defeq \{\pi : \mbox{$L_\Gmsc(\bfy_{\bfv_n})$
computable at the fusion center}\}.
\]
Note that the optimization in  (\ref{eq:optimal_reform}) is a 
function of the dependency graph $\Gmsc(\bfv_n)$, and that the 
optimal solution is attained by some policy. In general, the above 
optimization is NP-hard \cite{Anandkumar&etal:08INFOCOM}.

\subsection{Minimum energy data fusion: a lower bound}\label{sec:iid}
The following theorem  gives a lower bound on the minimum energy in 
(\ref{eq:optimal_reform}), given the joint dependency graph $\Gmsc $ 
and the path-loss exponent $\nu$. Let $\mbox{MST}(\bfv_n)$ be the 
Euclidean minimum spanning tree over a realization of sensor 
locations $\bfV_n=\bfv_n$.
\begin{theorem}[Lower bound on minimum energy expenditure]\label{thm:mst}
The following results hold:
\begin{enumerate}
\item the energy cost for the  optimal fusion policy $\pi^*$ in 
(\ref{eq:optimal_reform})   satisfies \beq \Etot(\pi^*(\bfv_n))\geq 
\Etot(\mst(\bfv_n)) \defeq 
 \sum_{e\in \smst(\bfv_n)} |e|^\nu 
 \label{eqn:lowerbndmst},
\eeq
\item
the lower bound (\ref{eqn:lowerbndmst}) is achieved (i.e., equality 
holds) when the observations are   independent under both 
hypotheses. In this case, the optimal   fusion policy $\pi^*$ 
aggregates data along $\mbox{DMST}(\bfv_n; v_1)$, the directed 
minimum spanning tree, with all the edges directed toward the fusion 
center $v_1$. Hence, the optimal fusion digraph $\Fmsc_{\pi^*}$ is 
the $\mbox{DMST}(\bfv_n; v_1)$.
% at its root and using edge-weight $|e|^\nu$ for
%$e\in \mbox{DMST}(\Vmsc; V_1)$.
\end{enumerate}
\end{theorem}

\begin{figure}[htb]
\begin{center}
\begin{psfrags}
\psfrag{FC}[l]{\small Fusion center} \psfrag{V0}[c]{\scriptsize
$v_1$} \psfrag{V1}[c]{\scriptsize $v_7$} \psfrag{V2}[c]{\scriptsize
$v_2$} \psfrag{V3}[c]{\scriptsize $v_3$} \psfrag{V4}[c]{\scriptsize
$v_4$} \psfrag{V5}[c]{\scriptsize $v_5$} \psfrag{V6}[c]{\scriptsize
$v_6$} \psfrag{q1}[c]{\scriptsize $q_1$} \psfrag{q2}[l]{\scriptsize
$q_2=L_2(y_2)+q_4+q_5$} \psfrag{q3}[c]{\scriptsize $q_3$}
\psfrag{q4}[c]{\scriptsize $q_4$} \psfrag{q5}[c]{\scriptsize $q_5$}
\psfrag{q6}[c]{\scriptsize $q_6$}
\includegraphics[width=2.5in]{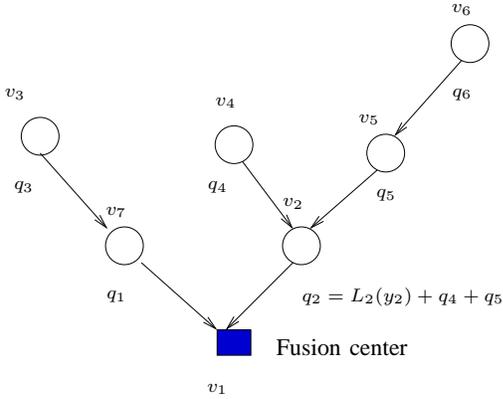}
\end{psfrags}
\end{center}
\caption{The optimal fusion graph DMST for   independent
observations.}\label{fig:MST}
\end{figure}

\bprf We  first prove part 2), for which we consider the case when 
observations are   independent, and the log-likelihood ratio is 
given by
\[
L_\Gmsc(\bfy_{\bfv_n})= \sum_{i\in \bfv_n} L_i(y_i),~~L_i(y_i)\defeq
\log\frac{f_{1,i}(y_i)}{f_{0,i}(y_i)},
\] where $f_{k,i}$ is the marginal pdf at node $i$ under $\Hc_k$.
Consider $\mbox{MST}(\bfv_n)$, whose links  minimize   
$\sum\limits_{e\in \mbox{\scriptsize Tree}(\bfv_n)} |e|^\nu$. It is 
easy to check that at the fusion center, the log-likelihood ratio 
can be computed using the following aggregation policy along the 
$\mbox{DMST}(\bfv_n;v_1)$ as illustrated in Fig.\ref{fig:MST}: each 
node $i$ computes the aggregated variable $q_i(\bfy_{\bfv_n})$ from 
its predecessor and sends it to its immediate successor. The 
variable $q_i$ is given by  the summation
\begin{equation}
q_i(\bfy_{\bfv_n}) \defeq \sum_{j\in \predecessor(i)} 
q_j(\bfy_{\bfv_n}) + L_i(y_i), \label{eq:agg}
\end{equation}
where $\predecessor(i)$ is the set of immediate predecessors of $i$
in $\mbox{DMST}(\bfv_n; v_1)$.

To show part 1), we note that any data-fusion policy must have each 
node transmit at least once and that the transmission must 
ultimately reach the fusion center. This implies that the fusion 
digraph must be connected with the fusion center and the DMST with 
edge-weight $|e|^\nu$ minimizes the total energy   under the above 
constraints. Hence, we have (\ref{eqn:lowerbndmst}).  \eprf

Note that the above lower bound in \eqref{eqn:lowerbndmst} is  
achievable when the measurements are   independent under both 
hypotheses. It is interesting to note that data correlations, in 
general, increase the energy consumption under the constraint of 
optimal inference performance since the log-likelihood ratio in 
\eqref{eqn:LLR} cannot be decomposed fully in terms of the 
individual node measurements.

\begin{figure*}[t]
\subfloat[a][Maximal cliques of dependency
graph]{\begin{minipage}{1.5in}
\begin{center}
\fbox{\includegraphics[height =
1in]{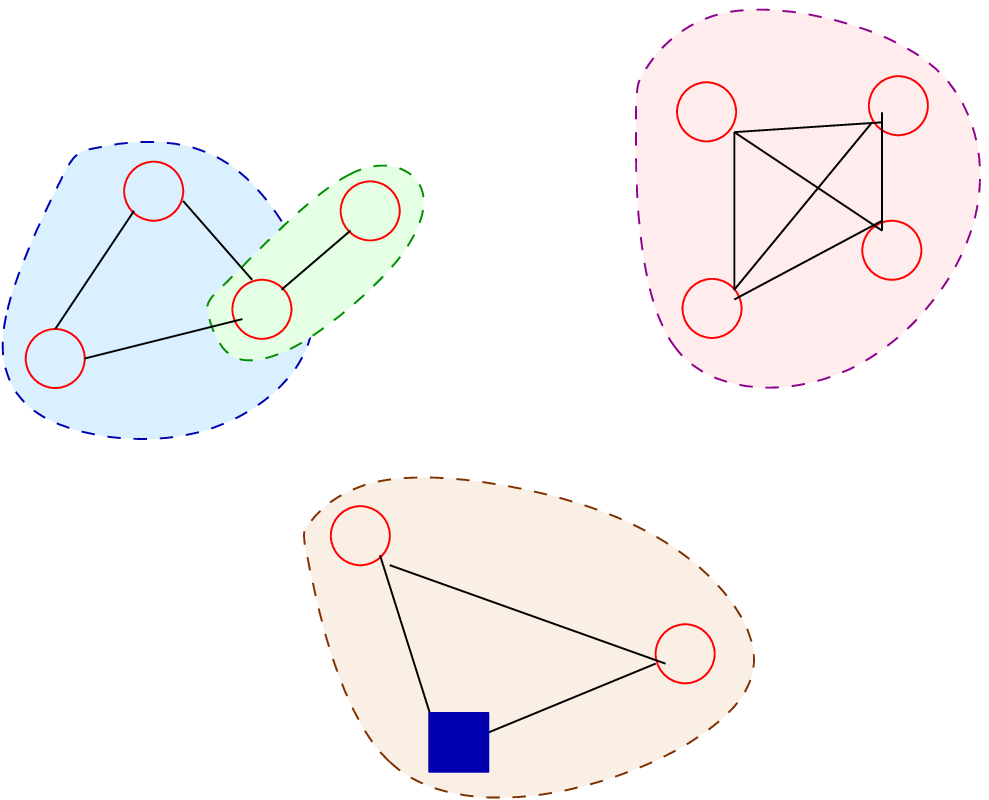}}\ec\end{minipage}}\hfil
\subfloat[b][Forwarding subgraph  computes clique
potentials]{\begin{minipage}{1.5in}\bc \fbox{\includegraphics[height
= 1in]{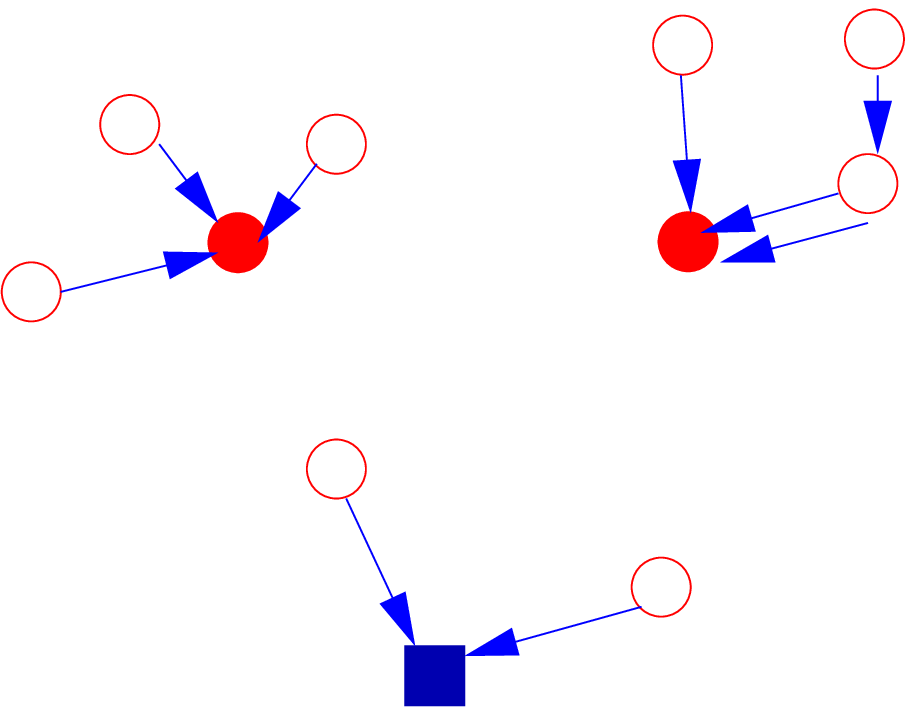}}\ec\end{minipage}}\hfil
\subfloat[c][Aggregation subgraph adds computed potentials
]{\begin{minipage}{1.5in}\bc\fbox{\includegraphics[height =
1in]{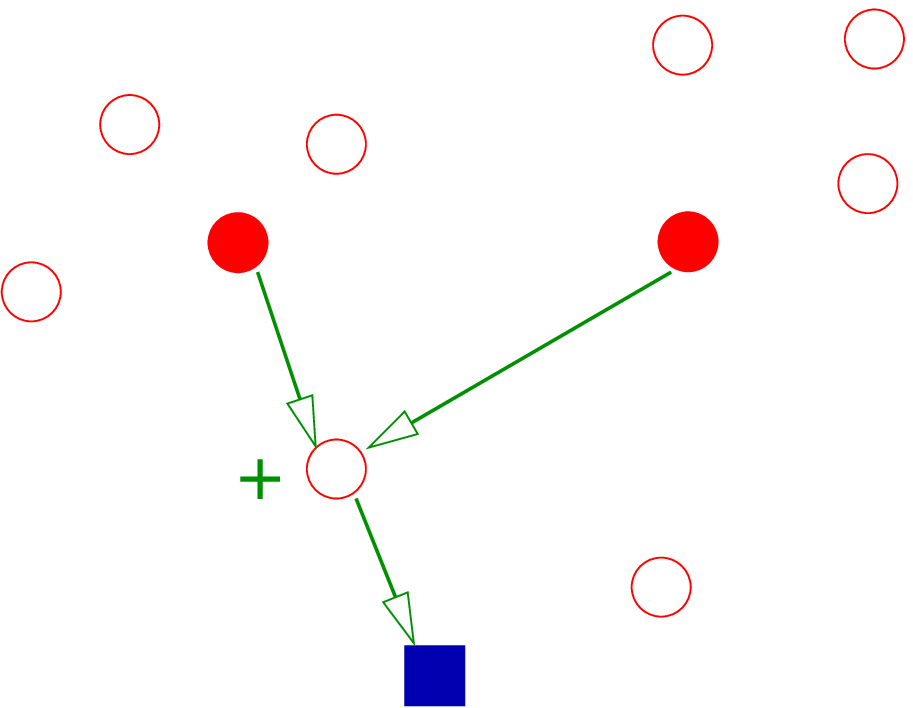}}\ec\end{minipage}}\hfil\subfloat[d][Legend]{\begin{minipage}{1.5in}\bc
\begin{psfrags}
 \psfrag{Edges of LCG}[l]{\scriptsize Forwarding subgraph (FG)}
 \psfrag{Edge in NNG}[l]{\scriptsize Dependency graph}
\psfrag{Edges of LAG}[l]{\scriptsize Aggregation graph
(AG)}\psfrag{Aggregator}[l]{\scriptsize Processor}\psfrag{Cluster
head or sink}[l]{\scriptsize  Fusion center}\fbox{
\includegraphics[height =1in,width=1.6in]{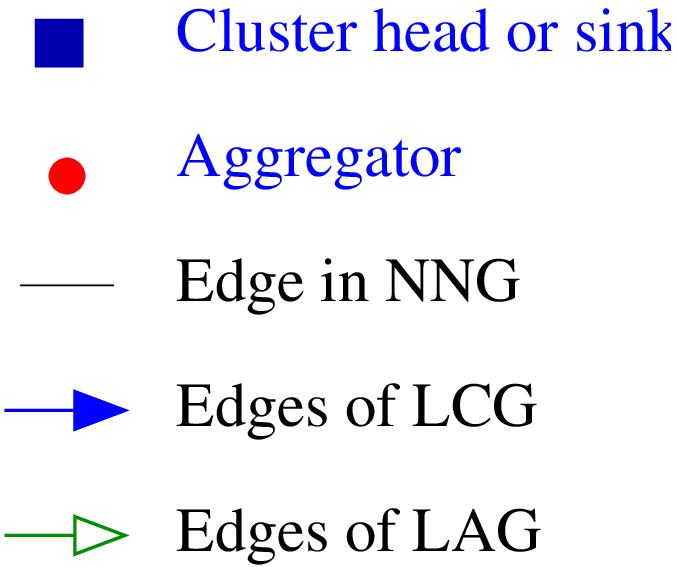}}\ep\ec\end{minipage}}\caption{Schematic of
dependency graph of Markov random field and stages of  data fusion.}
\label{fig:lcg_ag}
\end{figure*}

\subsection{Minimum energy data fusion: an upper bound}\label{sec:dfmrf}
We now  devise a suboptimal data-fusion policy which gives an upper 
bound on the optimal energy in (\ref{eq:optimal_reform})  for any 
given dependency graph $\Gmsc$ of the inference model. The 
suboptimal policy is referred to as Data Fusion on Markov Random 
Fields (DFMRF). It is a natural generalization of the MST 
aggregation policy,  described in Theorem \ref{thm:mst}, which is 
valid only for independent measurements.

We shall use Fig.~\ref{fig:lcg_ag} to illustrate the idea behind 
DFMRF. Recalling that the log-likelihood ratio for hypothesis 
testing of  Markov random fields is given by (\ref{eqn:LLR}), DFMRF  
  consists of two phases: 
\begin{enumerate}
\item  In the data forwarding phase, for each clique $c$ in the set of
maximal cliques $\Cmsc$ of the dependency graph $\Gmsc$, a {\em 
processor}, denoted by $\Proc(c)$, is chosen arbitrarily amongst the 
members of the clique $c$. Each node in clique $c$  (other than the 
processor itself)   forwards its raw data to $\Proc(c)$ via the 
shortest path using links in the network graph $\comm$. The 
processor $\Proc(c)$ computes the clique-potential function 
$\phi_c(\bfy_c)$ using the forwarded data.
\item  In the data-aggregation phase, processors compute the sum of the
clique potentials along $\mbox{DMST}(\bfv_n;v_1)$, the directed MST 
towards the fusion center, thereby delivering the log-likelihood 
ratio in \eqref{eqn:LLR} to the fusion center.
\end{enumerate} 

Hence, the fusion-policy digraph for  DFMRF  is   the union of the 
 subgraphs in the above two stages, \viz  forwarding subgraph 
($\mbox{FG}(\bfv_n)$) and   aggregation subgraph 
($\mbox{AG}(\bfv_n)$).  The total energy consumption of DFMRF is the 
sum of energies of the two subgraphs, given by \beqn\nn 
\Etot(\dfmrf(\bfv_n))&=& \sum_{c \in \Cc(\bfv_n)} \sum_{i \subset  
c} \spt(i,\Proc(c);\comm) \\&&+ \  \ \ 
\Etot(\mst(\bfv_n)),\label{eqn:dfmrf_orig}\eeqn where 
$\spt(i,j;\comm)$ denotes the energy consumption for the shortest 
path from $i$ to $j$ using the links in the network graph 
$\comm(\bfv_n)$ (set of feasible links for direct transmission). 
Recall that the network graph $\comm$ is different from the 
dependency graph $\Gmsc$ since the former deals with communication 
while the latter deals with data correlation.

%We can see  that aggregation along the DMST$(\Vmsc;\rootr)$ does not 
%deliver the LLR
%to the fusion center for a general MRF. This is because each function $\phi_c$ %has to
%%be computed before the summation, and it involves routing of raw
%aggregates raw measurements $\bfY_c$ at a common processor, and in 
%general, this 
% is not possible along the DMST.

For   independent measurements under either hypothesis, the maximal 
clique set $\Cmsc$ is trivially the set of vertices $\bfv_n$ itself 
and hence,   DFMRF  reduces to aggregation along the 
DMST$(\bfv_n;v_1)$, which is the optimal policy $\pi^*$ for   
independent observations. However, in general, DFMRF is not optimal. 
When the dependency graph $\Gmsc$ in \eqref{eqn:LLR} is the 
Euclidean  1-nearest neighbor  graph, we now show that the DFMRF has 
a constant approximation ratio with respect to the optimal 
data-fusion policy $\pi^*$ in (\ref{eq:optimal_reform}) for any 
arbitrary node placement.

\begin{theorem}[Approximation under 1-NNG dependency \cite{Anandkumar&Tong&Swami:07CISS}]
DFMRF is a 2-approximation fusion policy when the dependency graph 
$\Gmsc$ is the Euclidean 1-nearest neighbor graph for any fixed node 
set $\bfv_n\in \R^2$\beq \frac{\Etot(\dfmrf(\bfv_n))}{ 
\Etot(\pi^*(\bfv_n))} \leq 2. 
\label{eqn:approx_nng}\eeq\end{theorem}  

\bprf Since 1-NNG is acyclic,   the maximum clique size is 2. Hence, 
for DFMRF, the forwarding subgraph (FG) is  the 1-NNG with arbitrary 
directions on the edges. We have
\[\Etot(\lcg(\bfv_n))=\Etot(1\mbox{-}\nng(\bfv_n)) \leq \Etot(\mst(\bfv_n)).\]
Thus,
\begin{eqnarray}
\Etot(\mbox{DFMRF}(\bfv_n)) &=&
\Etot(\mbox{FG}(\bfv_n))+\Etot(\mbox{AG}(\bfv_n)),  \\ &\le 2&
\Etot(\mbox{MST}(\bfv_n)) \le 2 \Etot(\pi^*(\bfv_n)),  
\end{eqnarray}
where the last inequality comes from Theorem~\ref{thm:mst}. \eprf 

Note that the above result does not extend to general $k$-NNG 
dependency graphs $(k>1)$ for finite network size $n$. However, as 
the network size goes to infinity $(n\to \infty)$, we  show in 
Section \ref{sec:dfmrf_scaling} that a constant-factor approximation 
ratio is achieved by the DFMRF policy.

\section{Energy Scaling Laws}\label{sec:scaling}
We now establish the scaling laws for optimal and suboptimal fusion 
policies.  From the expression of average energy cost in 
(\ref{eqn:fusiondigraph}), we see that the scaling laws  rely on the 
law of large numbers (LLN) for stabilizing graph functionals. An 
overview of the LLN is provided in  Appendix \ref{sec:lln}.

We recall some notations and definitions used in this  section. $X_i 
\overset{i.i.d.}{\sim}\tau$, where $\tau$ is supported on $\Bc_1$, 
the unit square centered at the origin $\0$. The node location-set 
is $\Vmsc_n 
\defeq \sqrt{\frac{n}{\lambda}} (X_i)_{i=1}^n$ and the  limit is 
obtained by  letting $n\to \infty$ with fixed $\lambda>0$.

\subsection{Energy scaling for optimal fusion:
independent case}\label{sec:iid_scaling}

We first provide the scaling result for the case when the 
measurements are independent under either hypothesis. From Theorem 
\ref{thm:mst}, the optimal fusion policy minimizing the total energy 
consumption in \eqref{eq:optimal_reform} is given by aggregation 
along the directed minimum spanning tree. Hence, the energy scaling 
is obtained by the asymptotic analysis of the MST.

For the random node-location set $\Vmsc_n$, the average energy 
consumption of the optimal fusion policy for independent 
measurements is \beq \bar{\Etot}(\pi^*(\Vmsc_n)) = 
\bar{\Etot}(\mst(\Vmsc_n))= \frac{1}{n}\sum_{e \in \smst(\Vmsc_n)} 
|e|^\nu.\label{eqn:opt_iid}\eeq

Let $\constgd(\nu;\mst)$ be the constant arising in the asymptotic 
analysis of the MST edge lengths, given by \beq 
\constgd(\nu;\mst)\defeq\Ebb\Bigl[\sum_{e\in 
E(\0;\smst(\Pc_1\cup\{\0\}))}\frac{1}{2}|e|^\nu\Bigr],\label{eqn:constgd_def_mst}\eeq 
where  $\Pc_a$ is the homogeneous Poisson process of intensity 
$a>0$, and $E(\0;\mst(\Pc_1\cup\{\0\}))$ denotes the  set of edges 
incident to the origin in $\mst(\Pc_1\cup\{\0\})$. Hence, the above 
constant is half the expectation of the power-weighted edges 
incident to the origin in the minimum spanning tree over a 
homogeneous unit intensity Poisson process, and is discussed in 
Appendix \ref{sec:lln} in (\ref{eqn:constgd_derivation}). Although  
$\constgd(\nu;\mst)$ is  not available in closed form, we evaluate 
it through simulations in Section \ref{sec:num}.

We now provide the scaling result for the optimal fusion policy when 
the measurements are independent   based on the LLN for the MST 
obtained in \cite[Thm 2.3(ii)]{Penrose&Yukich:03AAP}.

\bt[Scaling for independent data \cite{Penrose&Yukich:03AAP}] 
\label{thm:iid_scaling} When the sensor measurements are independent 
under each hypothesis,  the limit of the average   energy 
consumption of the optimal fusion policy in (\ref{eqn:opt_iid}) is 
given by\beq \lim_{n \to \infty} 
\bar{\Etot}(\pi^*(\Vmsc_n))\overset{L^2}{=}\,\, 
\lambda^{-\frac{\nu}{2}}  \constgd(\nu;\mst)\int_{\Bc_{1}} \!\!\! 
\tau(x)^{1-\frac{\nu}{2}}dx.\label{eqn:iid_scaling} \eeq  \et

Hence, asymptotically the average energy consumption of optimal 
fusion is a constant   (independent of $n$)  in the mean-square 
sense for independent measurements. In contrast, forwarding all the 
raw data to the fusion center according to the shortest-path (SP) 
policy  has an unbounded average energy growing in the order of 
$\sqrt{n}$. Hence, significant energy savings are achieved through 
data fusion.

The scaling constant for average energy in (\ref{eqn:iid_scaling}) 
brings out the influence of several factors on energy consumption. 
It is inversely proportional to the node density $\lambda$. This is 
intuitive since placing the nodes with a higher density (\ie in a 
smaller area) decreases the average inter-node distances and hence, 
also the energy consumption. 

\begin{figure}[t]
\bc
\begin{psfrags}\psfrag{Nu}[c]{\scriptsize Path-loss exponent $\nu$}\psfrag{Uniform is 
Worst-Case}[c]{\scriptsize Unif. worst}\psfrag{Uniform is 
Optimal}[c]{\scriptsize Unif. best}\psfrag{Ratio in terms of uniform 
dist}[l]{\scriptsize $\int_{\Bc_{1}} \!\!\!\!\!\!
\tau(x)^{1-\frac{\nu}{2}}dx$} 
\includegraphics[width=2in,height=1.4in]{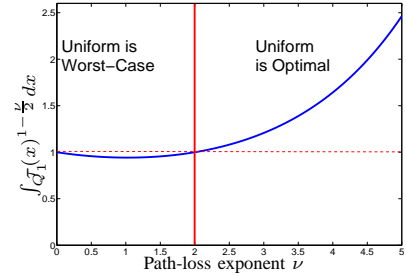}
\end{psfrags}\ec\caption{Ratio of energy consumption under node placement distribution
$\tau$ and uniform distribution as a function of path-loss exponent 
$\nu$. See (\ref{eqn:intg_lowerbnd}) and 
(\ref{eqn:intg_lowerbnd_nule2}).}\label{fig:mst_theo}\end{figure}

The node-placement pdf $\tau$ influences the asymptotic energy 
consumption through the term \[\int_{\Bc_{1}} \!\!\! 
\tau(x)^{1-\frac{\nu}{2}}dx.\] When the placement is uniform 
$(\tau\equiv1)$, the above term evaluates to unity.  Hence, the 
scaling constant in  (\ref{eqn:iid_scaling}) for uniform placement 
simplifies to
\[\lambda^{-\frac{\nu}{2}}  \constgd(\nu;\mst).\] The next theorem
shows  that the energy under uniform node placement $(\tau\equiv 1)$ 
optimizes the scaling limit in (\ref{eqn:iid_scaling})  when the 
path-loss exponent $\nu 
>2$.  Also, see Fig.\ref{fig:mst_theo}.

\bt[Minimum energy placement: independent 
case]\label{lemma:unif_mst} For any pdf $\tau$ supported on the unit 
square $\Bc_1$, we have \beqn\label{eqn:intg_lowerbnd}\int_{\Bc_{1}} 
\!\!\! \tau(x)^{1-\frac{\nu}{2}}dx&\geq& 1,\quad \forall\,\nu> 2,\\ 
\int_{\Bc_{1}} \!\!\! \tau(x)^{1-\frac{\nu}{2}}dx&\leq& 1,\quad 
\forall\,\nu\in [0,2).\label{eqn:intg_lowerbnd_nule2} \eeqn \et

\bprf We have the H\"older  inequality \beq \pl\!\! f_1 f_2 \!\! 
\pl_1\leq \pl\!\! f_1\!\!\pl_p \pl\!\! f_2\!\!\pl_q, \quad\forall p> 
1, q = \frac{p}{p-1},\label{eqn:holder}\eeq where for any positive 
function $f$,  
\[ \pl\!\!f\!\!\pl_p \defeq \Bigl(\int_{\Bc_1} f(x)^p dx\Bigr)^{\frac{1}{p}}.\]
When  $\nu> 2$, in (\ref{eqn:holder}),   substitute $f_1(x)$ with 
$\tau(x)^{\frac{1}{p}} $, $f_2(x)$ with $\tau(x)^{-\frac{1}{p}}$, 
and $p$ with $\frac{\nu}{\nu-2}\geq 1$  which ensures that $p>1$, to 
obtain (\ref{eqn:intg_lowerbnd}). 

For $\nu \in [0,2)$,  in (\ref{eqn:holder}), substitute $f_1(x)$ 
with $\tau(x)^{\frac{1}{p}} $, $f_2(x)$ with $1$, $p = 
\frac{2}{2-\nu}> 1$  to obtain (\ref{eqn:intg_lowerbnd_nule2}).\eprf

%
%\bprf Using the convexity of the function $g(x) = x^{1-
%\frac{\nu}{2}}$ for $\nu \geq 2$ over the range of $\tau$ we
%obtain via Jensen's inequality
%\[
%\int_{Q_1} (\tau(x))^{1 - \frac{\nu}{2}} dx \geq \Bigl( \int_{Q_1}
%\tau(x) dx\Bigr)^{1 - \frac{\nu}{2}} = 1.\] \eprf

The above result implies that, in the context of  i.i.d. node 
placements,  it is asymptotically energy-optimal to place the nodes 
uniformly when the path-loss exponent  $\nu >2$, which is the case 
for wireless transmissions. The intuitive reason   is as follows: 
without loss of generality, consider a clustered distribution in the 
unit square, where nodes are more likely to be placed near the 
origin. The MST over such a point set has many short edges, but a 
few very long edges, since a few nodes are placed near the boundary 
with finite probability. On the other hand, for uniform point sets, 
the edges of the MST are more likely to be all of similar lengths. 
Since for energy consumption, we have  power-weighted edge-lengths 
with  path-loss exponent $\nu 
>2$, long edges are penalized harshly, leading to higher energy consumption 
for clustered placement when compared with   uniform node placement.

%Moreover, we have that the integral in (\ref{eqn:intg_lowerbnd}) is 
%monotonically increasing in the path loss $\nu$. Hence, the uniform 
%placement has significantly lower energy consumption at higher path 
%loss coefficients.

\subsection{Energy scaling for optimal fusion: MRF
case}\label{sec:dfmrf_scaling}

We now evaluate the scaling laws for energy consumption of the DFMRF 
policy for a general Markov random field dependency  among  the 
sensor measurements.  The DFMRF aggregation policy involves the 
cliques of the dependency graph which arise from correlation between 
the sensor measurements. Recall that the total energy consumption of 
DFMRF in (\ref{eqn:dfmrf_orig}) for random sensor locations $\bfV_n$ 
is given by \beqn\nn \Etot(\dfmrf(\Vmsc_n))&=& \sum_{c \in 
\Cc(\Vmsc_n)} \sum_{i \subset  c} \spt(i,\Proc(c);\comm) 
\\&&+ \  \ \ \Etot(\mst(\Vmsc_n)),\label{eqn:dfmrf}\eeqn 
where $\spt(i,j;\comm)$ denotes the energy consumption for the 
shortest path between $i$ and $j$ using the links in the network 
graph $\comm(\Vmsc_n)$ (set of feasible links for direct 
transmission).  

We now additionally assume that the network graph $\comm(\Vmsc_n)$  
is a {\em local} $u$-energy spanner. In the literature 
\cite{Li:03WCMC}, a graph $\comm(\Vmsc_n)$  is called a $u$-energy 
spanner, for some constant $u>0$ called its {\em energy stretch 
factor}, when it satisfies \beq \label{eqn:uspanner_orig} 
\max_{i,j\in \Vmsc_n} 
\frac{\spt(i,j;\comm)}{\spt(i,j;\complete)}\leq u,\eeq where  
$\complete(\Vmsc_n)$ denotes the complete graph on $\bfV_n$.  In 
other words, the energy consumption between any two nodes is no 
worse than $u$-times the best possible value, \ie over the shortest 
path using links in the complete graph. Intuitively, the 
$u$-spanning property ensures that the network graph possesses 
sufficient set of communication links to ensure that the energy 
consumed in the forwarding stage is bounded. Examples of energy 
$u$-spanners include the Gabriel graph\footnote{The longest edge in 
Gabriel graph is $O(\sqrt{\log n})$, the same order as that of the 
MST \cite{Wan&Yi:07TranPDS}. Hence, the maximum power required at a 
node  to ensure $u$-energy spanning property is of the same order as 
that needed for critical connectivity.} (with stretch factor $u=1$ 
when the path-loss exponent $\nu\geq 2$), the Yao graph,  and its 
variations  \cite{Li:03WCMC}. In this paper, we only require a 
weaker version  of the above property that asymptotically there is 
at most $u$-energy stretch between the neighbors in the dependency 
graph\beq \label{eqn:uspanner} 
\limsup\limits_{n\to\infty}\max\limits_{(i,j)\in 
\Gmsc(\Vmsc_n)}\dfrac{\spt(i,j;\comm(\Vmsc_n))}{\spt(i,j;\complete(\Vmsc_n))}\leq 
u. \eeq   From (\ref{eqn:uspanner}), we have \beqn\nn 
\Etot(\lcg(\Vmsc_n)) &\leq& u \sum_{c \in \Cc(\Vmsc_n)} \sum_{i 
\subset c} \spt(i,\Proc(c);\complete),\nn
\\  &\leq&  u \sum_{c \in \Cc(\Vmsc_n)} \sum_{i \subset c}
|i,\Proc(c)|^\nu,\label{eqn:bound_proc}\eeqn where we use the 
property that the multihop shortest-path route from each node $i$ to 
$\Proc(c)$ consumes no more energy than the direct one-hop 
transmission. 

In the DFMRF policy, recall that  the processors are members of the 
respective cliques, \ie $\Proc(c)\subset c$, for each clique $c$ in 
the dependency graph. Hence, in (\ref{eqn:bound_proc}), only the 
edges of the processors of all the cliques  are included in the 
summation. This is upper bounded by the sum of all the 
power-weighted edges of the dependency graph $\dep(\Vmsc_n)$. Hence, 
we have \beq \Etot(\lcg(\Vmsc_n))\leq  u \sum_{e\in \dep(\Vmsc_n)} 
|e|^\nu\label{eqn:lcg_bound}.\eeq From (\ref{eqn:dfmrf}), for the 
total energy consumption of the DFMRF policy, we have the upper 
bound, \beq \Etot(\dfmrf(\Vmsc_n))\leq  u \sum_{e \in \dep(\Vmsc_n)} 
|e|^\nu+ \Etot(\mst(\Vmsc_n)).\label{eqn:lcg_lag_bound}\eeq The 
above bound allows us to draw upon the general methods of asymptotic 
analysis for graph functionals presented in 
\cite{Penrose&Yukich:03AAP,Penrose&Yukich:07Bern}.

From (\ref{eqn:lcg_lag_bound}), the DFMRF policy  scales whenever 
the right-hand side of (\ref{eqn:lcg_bound}) scales. By Theorem 
\ref{thm:iid_scaling},   the energy consumption for aggregation 
along the MST scales. Hence,  we only need to  establish the scaling 
behavior of the first term in (\ref{eqn:lcg_bound}).

%When we speak of sub-critical disc graph or continuum percolation
%\cite{Grimm:book}, we mean that  the disc radius is subcritical for
%all homogeneous Poisson point processes having intensity\[\tau \in [
%\inf_{x \in \Bmsc_1} \tau(x), \sup_{x \in \Bmsc_1} \tau(x) ],\]
%where $\Bmsc_1$ is the unit square. Intuitively, this means that the
%range of sensor dependencies is limited.

%\bl[$k$-NNG]\label{lemma:knng}The normalized sum of power-weighted
%edges  of the $k$-Euclidean nearest neighbor graph is a  stabilizing
%functional and satisfies bounded-moments condition
%(\ref{eqn:moment}).\el
%
%
%\bprf From \cite[Thm 2.4]{Penrose&Yukich:03AAP}.\eprf% See Appendix \ref{proof:knng}

%the cliques could be smaller than this : say Gaussian but that is fine: mention this.
%\bl[Sub-Critical Disk Graph]\label{lemma:disc}The normalized sum of
%power-weighted edges distances of the disc graph in the sub-critical
%region is a  stabilizing functional  and satisfies bounded-moments
%condition  (\ref{eqn:moment}).  \el
%
%\bprf  See Appendix \ref{proof:disc}.\eprf

We   now prove  scaling laws governing  the energy consumption of 
DFMRF and we also establish its asymptotic approximation ratio with 
respect to the optimal fusion policy. This in turn also establishes 
the scaling behavior of the optimal policy.

\bt[Scaling of DFMRF Policy]\label{thm:dfmrf_scaling} When the 
dependency graph $\dep$ of the sensor measurements is either  the 
$k$-nearest neighbor  or the disc graph, the  average energy  of 
DFMRF policy satisfies  \beqn&&\!\!\!\!\!\!\!\!\!\! \nn 
\limsup_{n\to \infty}\bar{\Etot}(\dfmrf(\Vmsc_n)) \\ \nn 
&&\!\!\!\!\!\!\!\!\!\! \overset{\as}{\leq} \limsup_{n\to 
\infty}\Bigl(\frac{1}{n}\sum_{e\in\sdep(\Vmsc_n)}  u \, |e|^\nu +
\bar{\Etot}(\mst(\Vmsc_n))\Bigr)\label{eqn:lcg_lag_bound_norm}\\
&&\!\!\!\!\!\!\!\! \overset{L^2}{=}\,\,    \frac{u}{2}\int_{\Bc_{1}}
\Ebb\Bigl[\!\!\!\!\sum_{j:(\0,j) \in
\sdep(\Pc_{\lambda\tau(x)}\cup\{\0\})}\!\!   |\0,j|^\nu   \Bigr]
\tau(x)dx\nn\\ &&+ \,
\lambda^{-\frac{\nu}{2}}\constgd(\nu;\mst)\int_{\Bc_{1}}
\tau(x)^{1-\frac{\nu}{2}}dx.\label{eqn:dfnngscalingclique_notscale}
\eeqn

\et

\bprf See Appendix \ref{proof:dfmrf_scaling}.\eprf

Hence, the above result establishes the scalability of the DFMRF
policy. In the theorem below, we use this result to prove the
scalability of the optimal fusion policy and establish asymptotic
upper and lower bounds on its average energy.

\bt[Scaling of Optimal Policy]\label{thm:opt_scaling} When the 
dependency graph $\dep$ is either  the $k$-nearest neighbor  or the 
disc graph,  the limit of the   average energy consumption of the 
optimal policy $\pi^*$ in \eqref{eq:optimal_reform} satisfies the 
upper bound \beq \limsup_{n\to 
\infty}\bar{\Etot}(\pi^*(\Vmsc_n))\overset{\as}{\leq}\limsup_{n\to 
\infty}\bar{\Etot}(\dfmrf(\Vmsc_n)), \eeq where the right-hand side 
satisfies the upper bound in 
(\ref{eqn:dfnngscalingclique_notscale}).  Also,  $\pi^*$ satisfies 
the lower bound given by the MST \beqn 
\!\!\!\!\!&&\!\!\!\!\!\!\!\!\!\! \!\! \liminf_{n\to 
\infty}\bar{\Etot}(\dfmrf(\Vmsc_n))\overset{\as}{\geq}\liminf_{n\to 
\infty}\bar{\Etot}(\pi^*(\Vmsc_n))\nn\\ 
\!\!\!\!\!&&\!\!\!\!\!\!\!\!\!\! \!\! \!\!\!\! \overset{\as}{\geq}  
\!\!\lim_{n \to \infty}\bar{\Etot}(\mst(\Vmsc_n)) \overset{L^2}{=} 
\lambda^{-\frac{\nu}{2}}\constgd(\nu;\mst)\int_{\Bc_{1}} \!\!\!\! 
\tau(x)^{1-\frac{\nu}{2}}dx.\label{eqn:lowerbndlimit}\eeqn \et

\bprf From (\ref{eqn:lowerbndmst}), the DFMRF and the optimal policy
satisfy the lower bound given by the MST.\eprf

Hence, the limiting average energy consumption for both the DFMRF 
policy and the optimal policy is strictly finite, and is bounded by 
(\ref{eqn:dfnngscalingclique_notscale}) and 
(\ref{eqn:lowerbndlimit}). These bounds also establish that the 
approximation ratio of the DFMRF policy is asymptotically bounded by 
a constant, as stated below. Define the constant $\approxratio:= 
\approxratio(u, \lambda, \tau, \nu)$, given by \beq 
\approxratio\defeq 1+\frac{u\displaystyle\int_{\Bc_{1}} 
\frac{1}{2}\Ebb\Bigl[\!\!\sum\limits_{j:(\0,j) \in 
\sdep(\Pc_{\lambda\tau(x)}\cup\{\0\})}\!\!   |\0,j|^\nu \Bigr] 
\tau(x)dx 
}{\lambda^{-\frac{\nu}{2}}\constgd(\nu;\mst)\displaystyle\int_{\Bc_{1}} 
\tau(x)^{1-\frac{\nu}{2}}dx} 
.\label{eqn:approxratio_notscale_def}\eeq

\bl[Approximation Ratio for DFMRF] The approximation ratio of
$\dfmrf$  is given by \beqn\nn&& \limsup_{n\to
\infty}\frac{\Etot(\dfmrf(\Vmsc_n))}{\Etot(\pi^*(\Vmsc_n))}\\
&&\overset{\as}{\leq}\,\, \limsup_{n\to
\infty}\frac{\Etot(\dfmrf(\Vmsc_n))}{\Etot(\mst(\Vmsc_n))}\,\,\overset{L^2}{=}\,\,\rho
\label{eqn:approxratio_notscale},\eeqn where $\approxratio$ is given
by (\ref{eqn:approxratio_notscale_def}).\el

\bprf Combine Theorem \ref{thm:dfmrf_scaling} and Theorem
\ref{thm:opt_scaling}.\eprf

We  further simplify the above results for the $k$-nearest neighbor 
dependency graph  in the corollary below by exploiting its scale 
invariance.  The results are expected to hold for  other {\em 
scale-invariant} Euclidean  stabilizing  graphs as well. The edges   
of a scale-invariant graph are invariant under a change of scale, or 
put differently, $\Gmsc$ is scale invariant if scalar multiplication 
by any positive constant $\alpha$  from $\Gmsc(\Vmsc_n)$ to 
$\Gmsc(\alpha \Vmsc_n)$ induces a graph isomorphism for all node 
sets $\Vmsc_n$.

Along the lines of (\ref{eqn:constgd_def_mst}), let
$\constgd(\nu;k\mbox{-}\nng)$ be the constant arising in the
asymptotic analysis of the $k$-NNG  edge lengths, that is \beq
\constgd(\nu;k\mbox{-}\nng)\defeq\Ebb\Bigl[\sum_{j:(\0,j)\in
k\mbox{\tiny -}\snng(\Pc_1\cup\{\0\})}\frac{1}{2}
|\0,j|^\nu\Bigr].\label{eqn:constgd_def}\eeq

\begin{corollary}[$k$-NNG Dependency Graph]\label{cor:dfmrf}We obtain a simplification of 
Theorem \ref{thm:dfmrf_scaling} and \ref{thm:opt_scaling} for 
average energy consumption, namely 
\beqn\!\!\!\!&&\!\!\!\!\!\!\!\!\!\!\limsup_{n\to 
\infty}\bar{\Etot}(\pi^*(\Vmsc_n))\overset{\as}{\leq}\limsup_{n\to 
\infty}\bar{\Etot}(\dfmrf(\Vmsc_n))\nn\\ 
\!\!\!\!&&\!\!\!\!\!\!\!\!\!\! \overset{\as}{\leq} \limsup_{n\to 
\infty}\Bigl(\frac{1}{n}\sum_{e\in\sdep(\Vmsc_n)}  u \, |e|^\nu + 
\bar{\Etot}(\mst(\Vmsc_n))\Bigr)\nn\\  \!\!\!\!&&\!\!\!\!\!\!\!\! 
\overset{L^2}{=}  \lambda^{-\frac{\nu}{2}} [ 
u\,\constgd(\nu;k\mbox{-}\nng)+\constgd(\nu;\mst)]\int_{\Bc_{1}} 
\!\!\!\! 
\tau(x)^{1-\frac{\nu}{2}}dx.\label{eqn:dfnngscalingclique}\eeqn The 
approximation ratio of $\dfmrf$  satisfies  \beqn\nn \limsup_{n\to 
\infty}\frac{\Etot(\dfmrf(\Vmsc_n))}{\Etot(\pi^*(\Vmsc_n))}\!\!\!\!&\overset{\as}{\leq}& 
\!\!\!\! \limsup_{n\to
\infty}\frac{\Etot(\dfmrf(\Vmsc_n))}{\Etot(\mst(\Vmsc_n))}\\
\!\!\!\!&\overset{L^2}{=}& \!\!\!\! 
\Bigl(1+u\frac{\constgd(\nu;k\mbox{-}\nng) 
}{\constgd(\nu;\mst)}\Bigr).\label{eqn:approxratio}\eeqn\end{corollary}

\bprf This follows from \cite[Thm 2.2]{Penrose&Yukich:03AAP}.\eprf

Hence, the expressions for the energy scaling bounds and the 
approximation ratio are further  simplified when the dependency 
graph is the $k$-nearest neighbor graph.  A special case of this 
scaling result for the 1-nearest-neighbor dependency under uniform  
node placement was proven in \cite[Thm 
2]{Anandkumar&Tong&Swami:08SP}.

It is interesting to note that the approximation factor for  the 
$k$-NNG dependency graph in (\ref{eqn:approxratio}) is  independent 
of the node placement pdf $\tau$ and node density $\lambda$.  Hence, 
DFMRF has the same efficiency relative to the optimal policy under 
different node placements. The results of Theorem 
\ref{lemma:unif_mst} on the optimality of the uniform node placement 
are also  applicable here, but for the lower and upper bounds on  
energy consumption. We formally state it below.

\bt[Minimum energy bounds for 
$k$-NNG]\label{lemma:uniform_knng}Uniform node placement  $(\tau 
\equiv 1)$ minimizes the asymptotic lower  and upper bounds  on  
average energy consumption  in (\ref{eqn:lowerbndlimit}) and 
(\ref{eqn:dfnngscalingclique}) for   the optimal policy under the 
$k$-NNG dependency graph  over all i.i.d. node placement  pdfs 
$\tau$.\et 

\bprf From Theorem \ref{lemma:unif_mst} and 
(\ref{eqn:dfnngscalingclique}).\eprf

We also prove the optimality of uniform node-placement distribution  
under the disc-dependency graph, but over a limited set of node 
placement pdfs $\tau$.

\bt[Minimum energy bounds for disc 
graph]\label{lemma:uniform_disc}Uniform node placement  $(\tau 
\equiv 1)$ minimizes the asymptotic lower and upper bounds  on the  
average energy consumption  in (\ref{eqn:lowerbndlimit}) and 
(\ref{eqn:dfnngscalingclique}) for the optimal fusion policy under 
the disc dependency graph over all i.i.d. node-placement  pdfs 
$\tau$ satisfying the lower bound\beq \tau(x) > 
\frac{1}{\lambda},\quad\forall x \in 
\Bc_1,\label{eqn:kappa_bound}\eeq where $\lambda > 1$ is the (fixed) 
node placement density.\et

\bprf We use the fact that for the disc graph  $\Gmsc$ with a fixed 
radius, more edges are added as we scale down  the area. Hence,   
for Poisson processes with intensities $\lambda_1 > \lambda_2>0$,\[  
\Ebb\Bigl[\!\!\!\!\!\!\sum_{j:(\0,j) \in 
\sdep(\Pc_{\lambda_1}\cup\{\0\})}\!\!\!\!   |\0,j|^\nu   \Bigr] \geq 
\Ebb\Bigl[\!\!\!\!\!\!\sum_{j:(\0,j) \in 
\sdep(\Pc_{\lambda_2}\cup\{\0\})}\!\!\!\!   |\0,j|^\nu   
\Bigr]\left[\frac{\lambda_2}{\lambda_1}\right]^{\frac{\nu}{2}},\] 
where the right-hand side is obtained by merely rescaling the edges 
present under the Poisson process at intensity $\lambda_2$. Since, 
new edges are added under 
 the Poisson process at $\lambda_1$, the above expression is an inequality, unlike 
the case of $k$-NNG where  the edge set is invariant under scaling. 
Substituting $\lambda_1$ with  $\lambda \tau(x)$, and $ \lambda_2$ 
by $1$ under the condition that $\lambda \tau(x)> 1$, $\forall x \in 
\Bc_1$,  we have\beqn\nn  && 
\int_{\Bc_1}\Ebb\Bigl[\!\!\!\!\!\!\sum_{j:(\0,j) \in 
\sdep(\Pc_{\lambda\tau(x)}\cup\{\0\})}\!\!\!\!   |\0,j|^\nu   \Bigr] 
\tau(x)dx\\ &\geq& \lambda^{-\frac{\nu}{2}} 
\Ebb\Bigl[\!\!\!\!\!\!\sum_{j:(\0,j) \in 
\sdep(\Pc_{1}\cup\{\0\})}\!\!\!\!   |\0,j|^\nu   
\Bigr]\int_{\Bc_1}\tau(x)^{1-\frac{\nu}{2}}dx,\nn\\ &\geq& 
\lambda^{-\frac{\nu}{2}} \Ebb\Bigl[\!\!\!\!\!\!\sum_{j:(\0,j) \in 
\sdep(\Pc_{1}\cup\{\0\})}\!\!\!\!   |\0,j|^\nu   \Bigr]\nn,\quad 
\nu>2.\eeqn  \eprf

Hence, uniform node placement is optimal in terms of the energy 
scaling bounds under the disc dependency graph if we restrict to 
pdfs $\tau$ satisfying (\ref{eqn:kappa_bound}). 

We have so far established the finite scaling of the average energy 
when the dependency graph describing the correlations among the 
sensor observations is either the $k$-NNG or the disc graph with 
finite radius. However, we cannot expect finite energy scaling under 
any general dependency graph. For instance, when the dependency 
graph is the     complete graph, the log-likelihood ratio in 
(\ref{eqn:LLR}) is   a function of only one clique containing all 
the nodes. In this case, the optimal policy in 
\eqref{eq:optimal_reform}   consists of  a unique  processor chosen 
optimally, to which all the other nodes forward their raw data along 
shortest paths, and the processor then forwards the value of the 
computed log-likelihood ratio to the fusion center. Hence, for the 
complete dependency graph,   the optimal fusion policy reduces to a 
version of the shortest-path (SP) routing, where the average energy 
consumption grows as $\sqrt{n}$ and does not scale with $n$.   

%Intuitively,  (\ref{eqn:kappa_bound}) guarantees that all nodes are 
%not clustered at a single point. The bound is inversely proportional 
%to node density (or directly proportional to area of placement). 
%Hence, in a large area, we are required  to ``spread" the 
%probability of placement among all the points.

\begin{figure*}[t]
\subfloat[a][Avg. energy vs. no. of nodes,
$\nu=2$.]{\label{fig:energy}\begin{minipage}{2.3in} \begin{center}
 \begin{psfrags}
\psfrag{Number of nodes}[c]{\scriptsize Number of nodes $n$}
\psfrag{Average energy}[c]{\scriptsize Avg. energy per node}
 \psfrag{1 nearest neighbor dependency}[l]{\scriptsize 1-NNG: DFMRF}\psfrag{3
nearest neighbor dependency}[l]{\scriptsize 3-NNG: DFMRF}\psfrag{2
nearest neighbor dependency}[l]{\scriptsize 2-NNG: DFMRF}
\psfrag{shortest path}[l]{\scriptsize No Fusion: SPR}
\psfrag{Uncorrelated data}[l]{\scriptsize 0-NNG: MST}
\includegraphics[width=2.1in,height=1.6in]{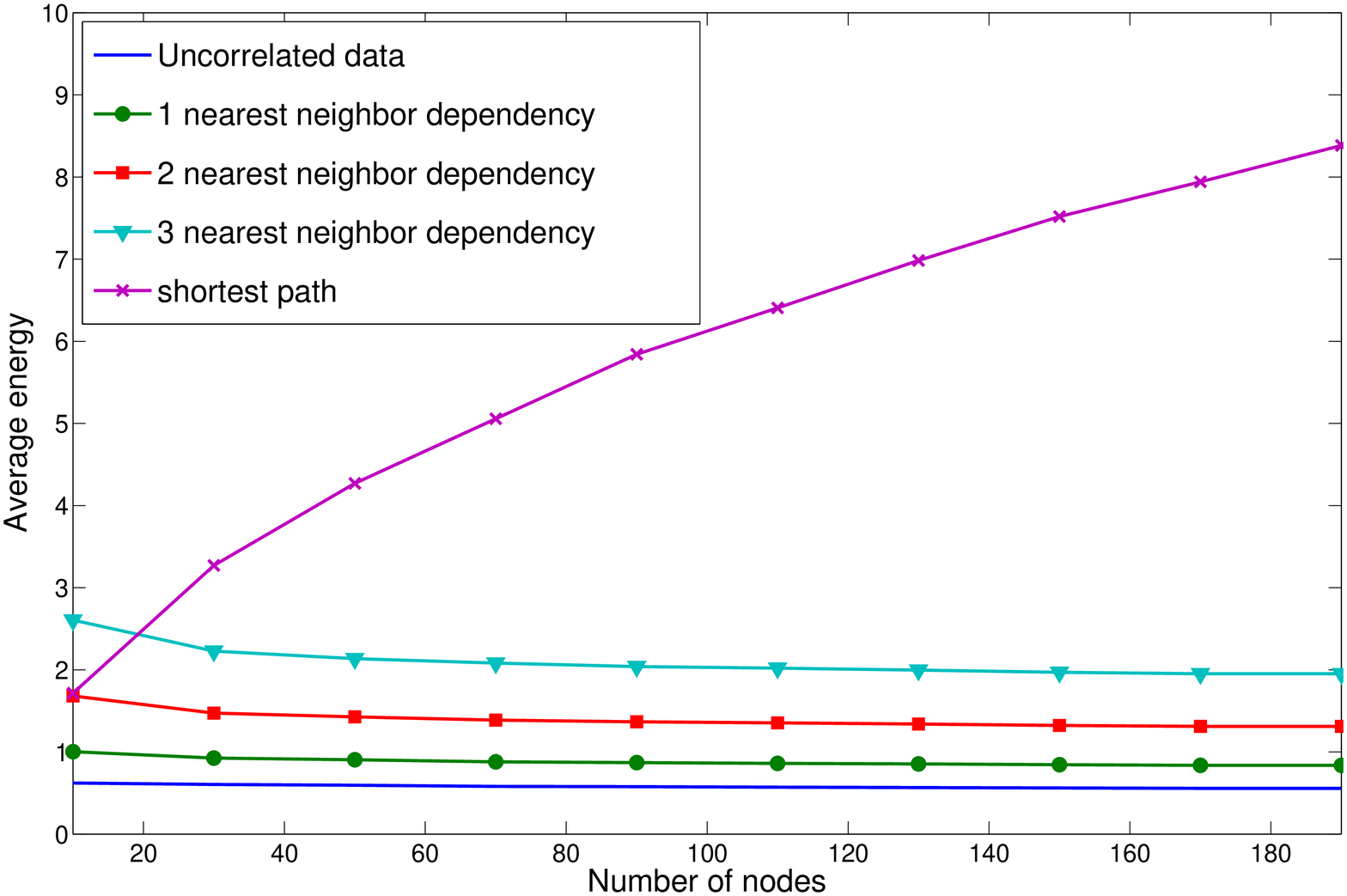}
\end{psfrags}
\end{center}\end{minipage}}
\subfloat[b][Approx. ratio vs. no. of nodes,
$\nu=2$.]{\label{fig:ratio_num}\begin{minipage}{2.3in}
  \begin{center}
 \begin{psfrags}
\psfrag{Number of nodes}[c]{\scriptsize Number of nodes $n$}
\psfrag{Approximation ratio}[c]{\scriptsize Approx. ratio for DFMRF}
 \psfrag{1 nearest neighbor dependency}[l]{\scriptsize 1-NNG dependency}\psfrag{3
nearest neighbor dependency}[l]{\scriptsize 3-NNG
dependency}\psfrag{2 nearest neighbor dependency}[l]{\scriptsize
2-NNG  dependency}   \psfrag{Uncorrelated data}[l]{\scriptsize No
correlation}
\includegraphics[width=2.1in,height=1.6in]{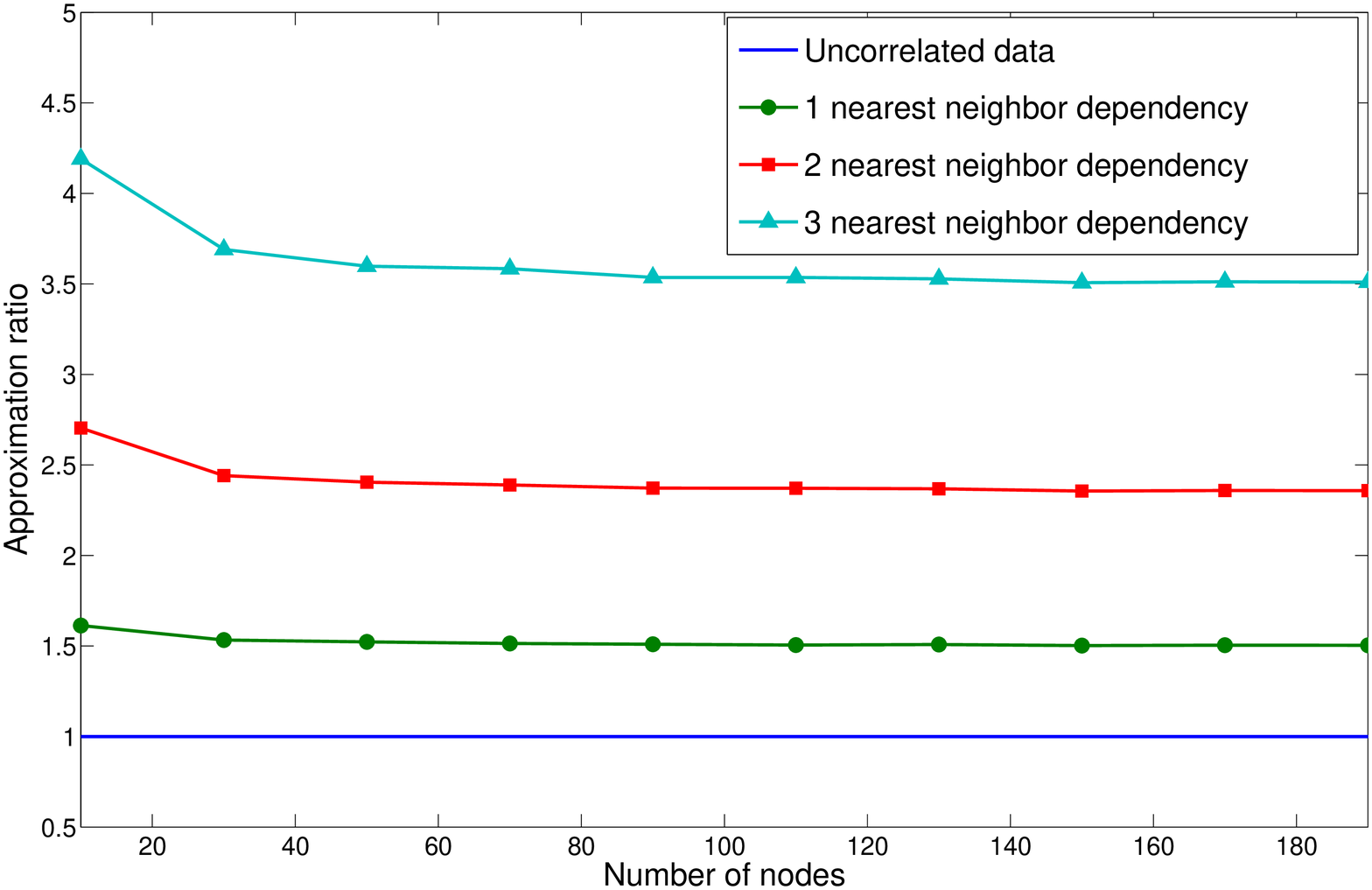}
\end{psfrags}
\end{center}\end{minipage}}
\subfloat[b][Approx. ratio vs. path-loss,
$n=190$.]{\label{fig:ratio_nu}\begin{minipage}{2.1in}
  \begin{center}
 \begin{psfrags}
\psfrag{Path Loss Nu}[c]{\scriptsize Path-loss exponent $\nu$}
\psfrag{Approximation ratio}[c]{\scriptsize Approx. ratio for DFMRF}  
\psfrag{1 nearest neighbor dependency}[l]{\scriptsize 1-NNG 
dependency}\psfrag{3 nearest neighbor dependency}[l]{\scriptsize 
3-NNG dependency}\psfrag{2 nearest neighbor 
dependency}[l]{\scriptsize 2-NNG  dependency}   \psfrag{Uncorrelated 
data}[l]{\scriptsize No correlation}
\includegraphics[width=2.1in,height=1.6in]{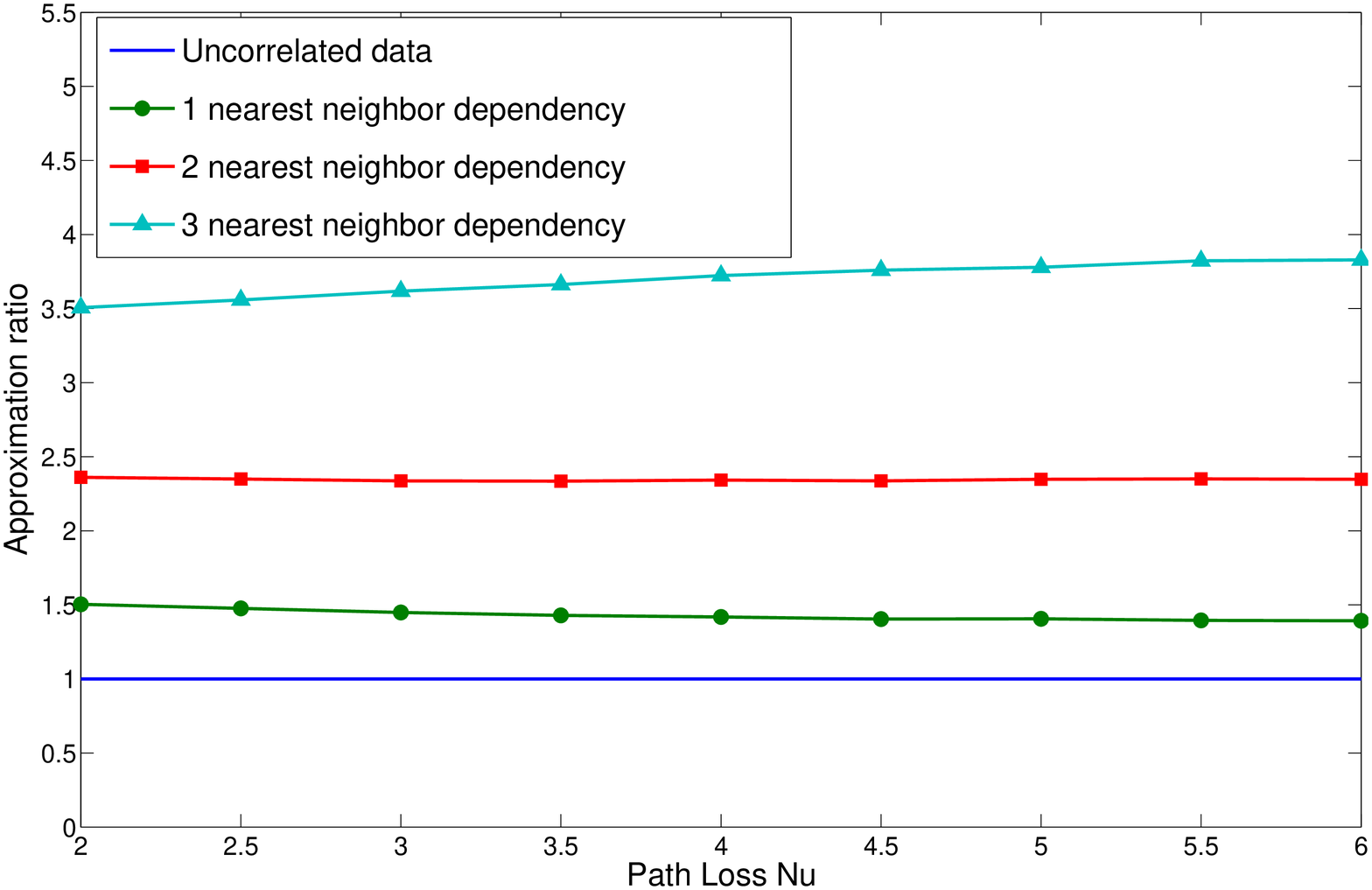}
\end{psfrags}
\end{center}\end{minipage}}
\caption{Average   energy consumption for DFMRF policy and 
shortest-path routing for uniform node distribution and $k$-NNG 
dependency  over 500 runs.   Node density $\lambda=1$. See Corollary 
\ref{cor:dfmrf}.}\label{fig:knng}
 \end{figure*}

\begin{figure*}[t]
\subfloat[a][Disk graph, $\nu=2$, uniform 
$(\tau\equiv1)$.]{\label{fig:energy_disk}\begin{minipage}{2.3in}
\begin{center}
 \begin{psfrags}
\psfrag{Number of nodes}[c]{\scriptsize Number of nodes $n$}
\psfrag{Average energy}[c]{\scriptsize Avg. energy for DFMRF}
 \psfrag{delta=d1}[l]{\scriptsize $\delta=0.3$} \psfrag{delta=d2}[l]{\scriptsize $\delta=0.6$}
 \psfrag{delta=d3}[l]{\scriptsize $\delta=0.9$}
\psfrag{Uncorrelated data}[l]{\scriptsize $\delta=0$}
\includegraphics[width=2.1in,height=1.6in]{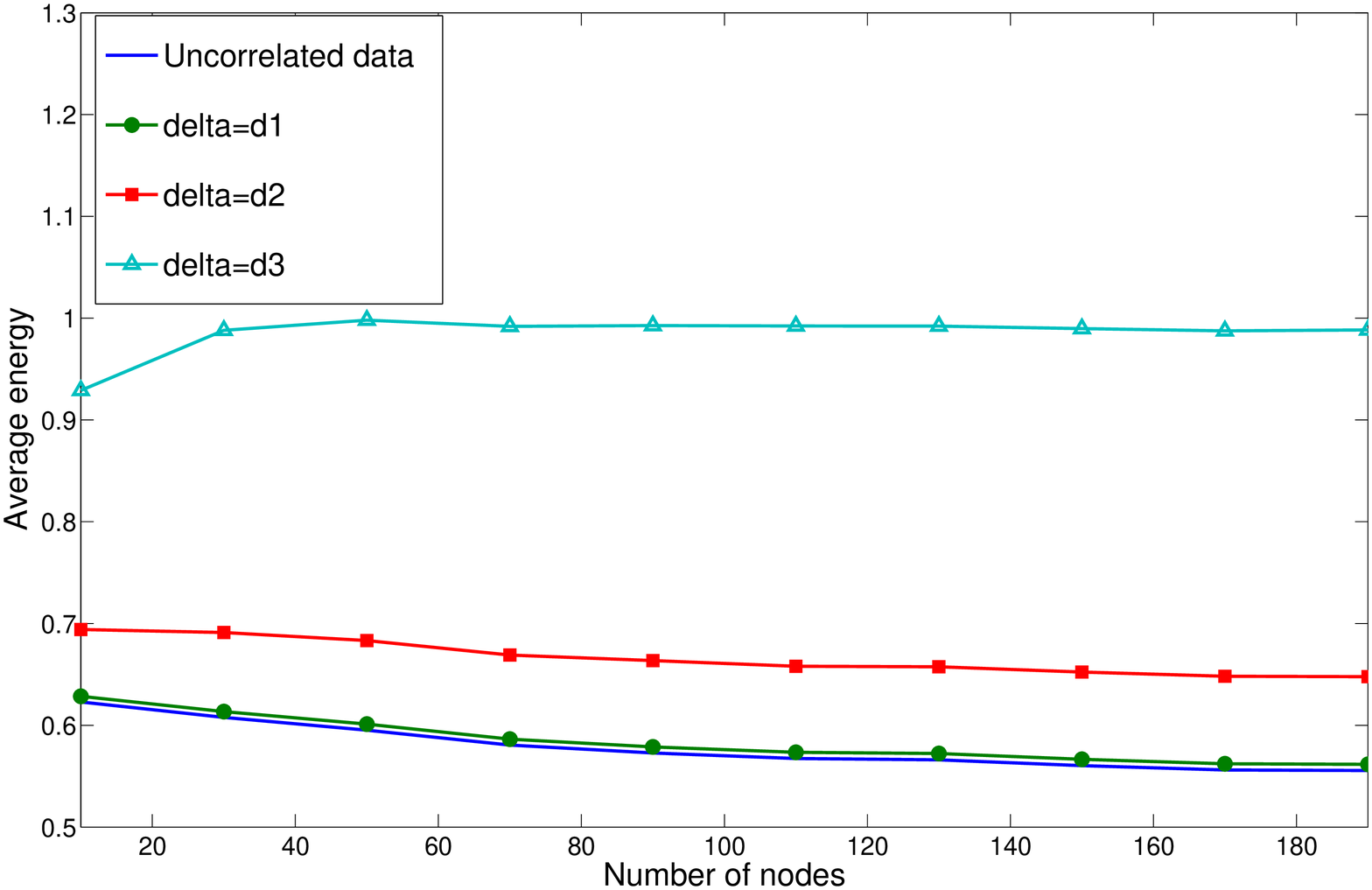}
\end{psfrags}
\end{center}\end{minipage}} \subfloat[b][Avg. energy vs. path loss, $\delta=0$, $n=190$.]
{\label{fig:mst_kappa}\begin{minipage}{2.1in}
\begin{center}\bp\psfrag{Nu}[c]{\scriptsize Path-loss exponent
$\nu$}\psfrag{Uniform}[l]{\scriptsize Uniform: 
$a=0$}\psfrag{Clustered}[l]{\scriptsize Clustered: 
$a=5$}\psfrag{Spread out}[l]{\scriptsize Spread out: 
$a=-5$}\psfrag{Avg. energy}[c]{\scriptsize Avg. Energy for 
DFMRF}\includegraphics[width=2.1in,height=1.6in]{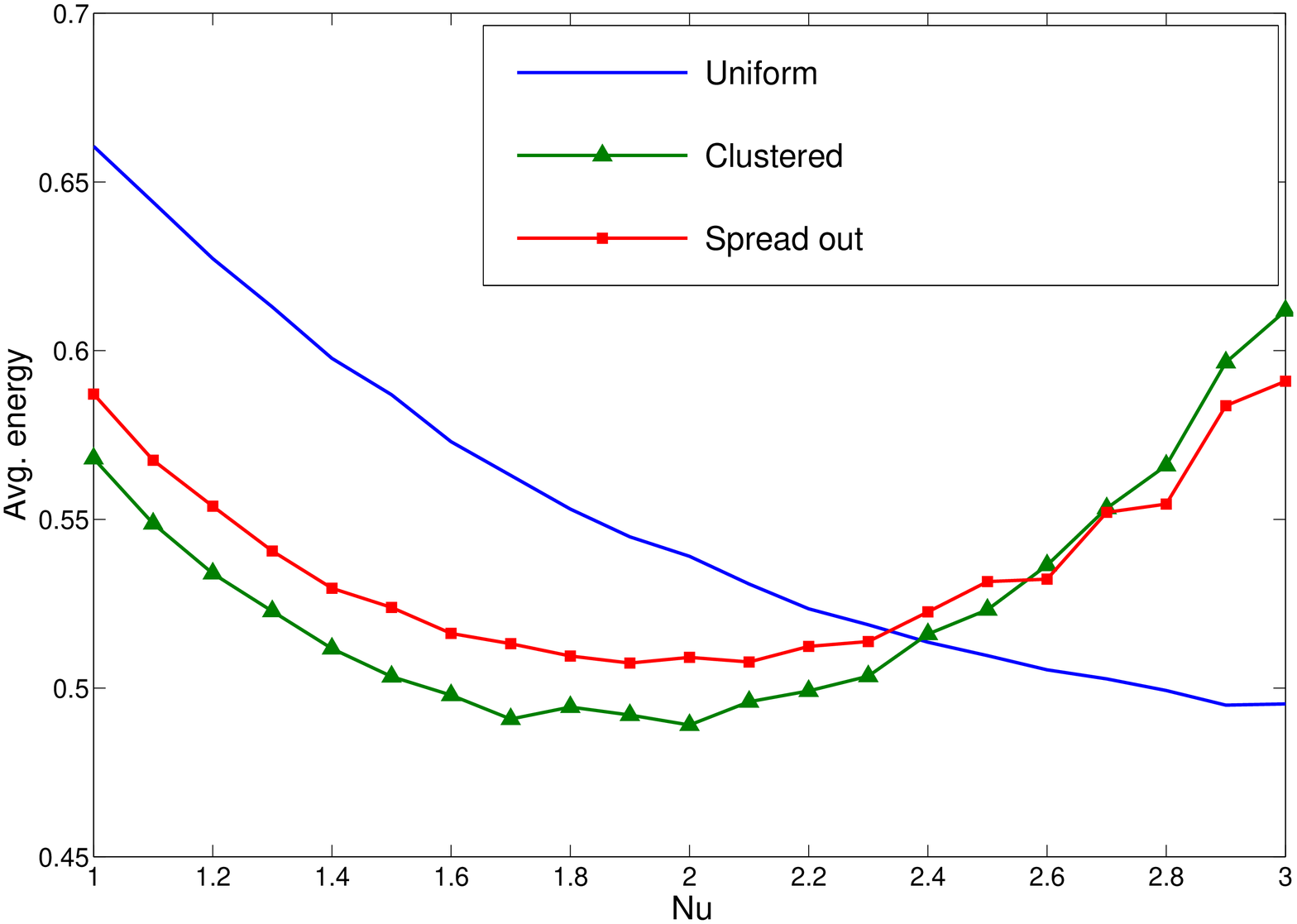}\ep
\end{center}\end{minipage}}\subfloat[c][Avg energy vs. disk radius,
$\nu=4$.]{\label{fig:disc_kappa_nu4}\begin{minipage}{2.3in}
\begin{center}
 \begin{psfrags}
\psfrag{Disk radius}[c]{\scriptsize Disk Radius $\delta$} 
\psfrag{Average energy}[c]{\scriptsize Avg. energy for DFMRF} 
\psfrag{Uniform: a=0}[l]{\scriptsize Uniform: 
$a=0$}\psfrag{Clustered: a=10}[l]{\scriptsize Clustered: $a=5$} 
\psfrag{Spread out: a=-10}[l]{\scriptsize Spread out: $a=-5$}
\includegraphics[width=2.1in,height=1.6in]{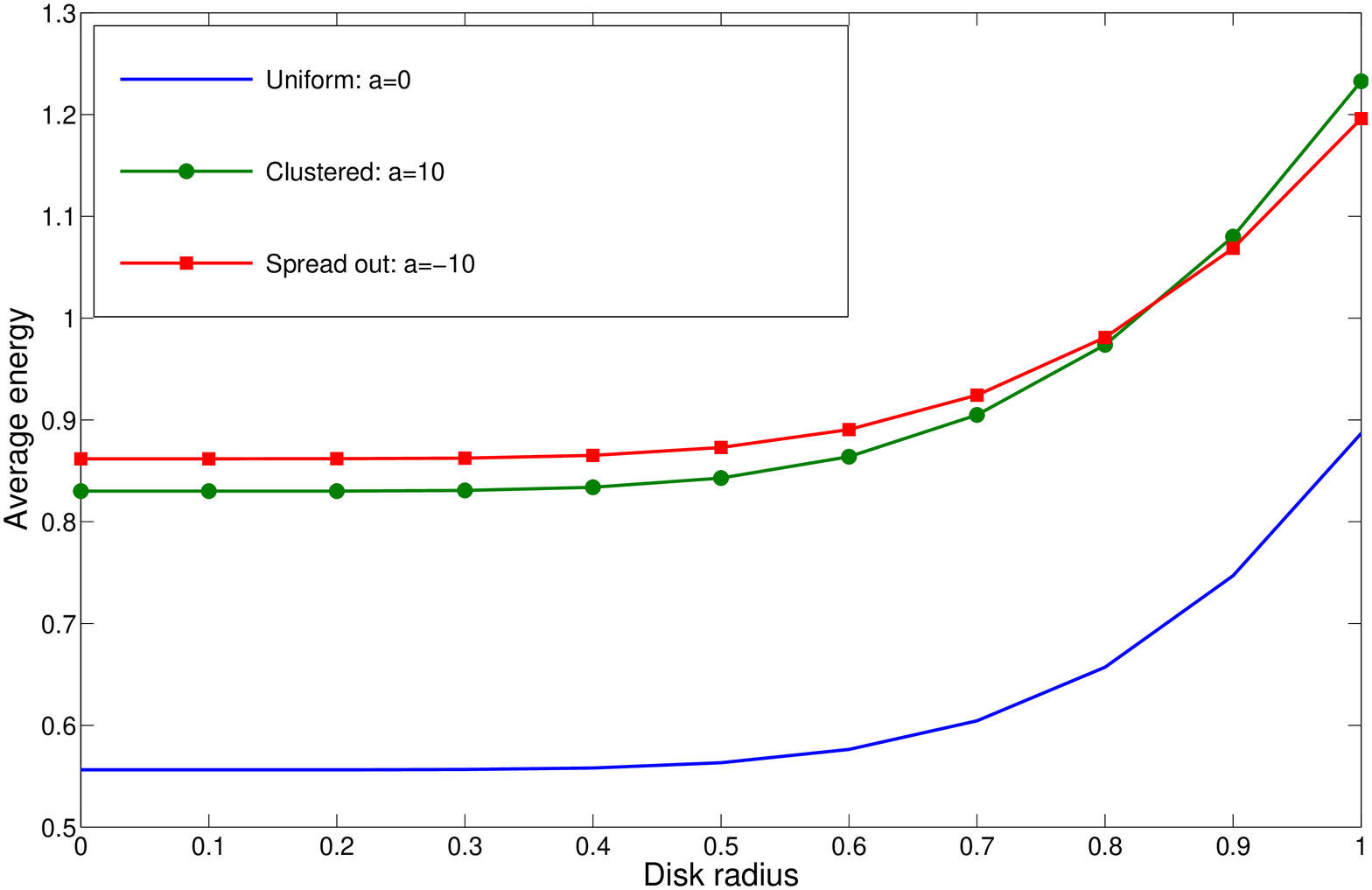}
\end{psfrags}
\end{center}\end{minipage}}
\caption{Average   energy consumption for DFMRF  policy over 500 
runs for node-placement pdfs shown in Fig.\ref{fig:points_kappa} 
under disc-dependency graph with radius $\delta$.  Node density 
$\lambda=1$. See Theorem 
\ref{thm:dfmrf_scaling}.}\label{fig:nu_kappa}
 \end{figure*}

\section{Numerical Illustrations}\label{sec:num}

%The results in the previous sections (Theorem
%\ref{thm:dfmrf_scaling} and Corollary \ref{cor:dfmrf}) guarantee
%finite average energy scaling for the DFMRF fusion policy and also
%provide asymptotic bounds and the approximation ratio. In this
%section, we numerically evaluate these quantities through
%simulations under different node placement distributions and
%dependency graphs.Of the $n$ nodes, we uniformly pick one of them
%as the fusion center.

As described in Section \ref{sec:location}, $n$ nodes are placed in 
area $\frac{n}{\lambda}$ and one of them is randomly chosen as the 
fusion center.  We conduct 500 independent simulation runs and 
average the results.  We fix node density $\lambda=1$. We plot 
results for two cases of dependency graph, \viz the $k$-nearest 
neighbor graph and the disc graph with a fixed radius $\delta$.

In Fig.\ref{fig:knng}, we plot the simulation results for the 
$k$-nearest neighbor dependency graph and uniform node placement. 
Recall   in Corollary \ref{cor:dfmrf},  we established  that the 
average energy consumption of  the DFMRF policy in 
(\ref{eqn:dfnngscalingclique}) is finite and bounded for asymptotic 
networks under $k$-NNG dependency. On the other hand, we predicted 
in  Section \ref{sec:intro_scalable} that the average energy under 
no aggregation (SP policy) increases without bound with the network 
size. The results in  Fig.\ref{fig:energy} agree with our theory and 
we note that the convergence to asymptotic values is quick, and 
occurs in networks with as little as $30$ nodes.  We also see that   
the energy for DFMRF policy increases with the number of neighbors 
$k$ in the dependency graph since the graph has more edges leading 
to computation of a more complex likelihood ratio by the DFMRF 
policy. 

We plot the approximation ratio of the DFMRF policy for $k$-NNG in 
(\ref{eqn:approxratio}) against the number of nodes  in 
Fig.\ref{fig:ratio_num} and  against the path-loss exponent $\nu$ in 
Fig.\ref{fig:ratio_nu}. As established by Corollary \ref{cor:dfmrf}, 
the approximation ratio is a constant for large networks, and we 
find a quick convergence to this value in Fig.\ref{fig:ratio_num} as 
we increase the network size. In Fig.\ref{fig:ratio_nu},  we also 
find that the approximation ratio is fairly insensitive  with 
respect to the path-loss exponent $\nu$.  

In Fig.\ref{fig:energy_disk}, we plot the average energy consumption 
of DFMRF in (\ref{eqn:dfnngscalingclique_notscale}) under uniform 
node placement and the disc dependency graph with radius $\delta$. 
The   average energy is bounded, as established by Theorem 
\ref{thm:dfmrf_scaling}. As in the $k$-NNG case, on increasing the 
network size, there is a quick convergence to the asymptotic values.  
Moreover, as expected, energy consumption increases with the radius  
$\delta$ of the disc graph since there are more edges. Note that the 
energy consumption at $\delta=0$ and $\delta=0.3$ are nearly the 
same, since at $\delta=0.3$, the disc graph is still very sparse, 
and hence,  the energy consumed in the forwarding stage   of the 
likelihood-ratio computation is small.

 \begin{figure}[hbt]\subfloat[a][Uniform $a\to 0$.]{\begin{minipage}{1in}
 \bc\includegraphics[width=1in,height=1in]{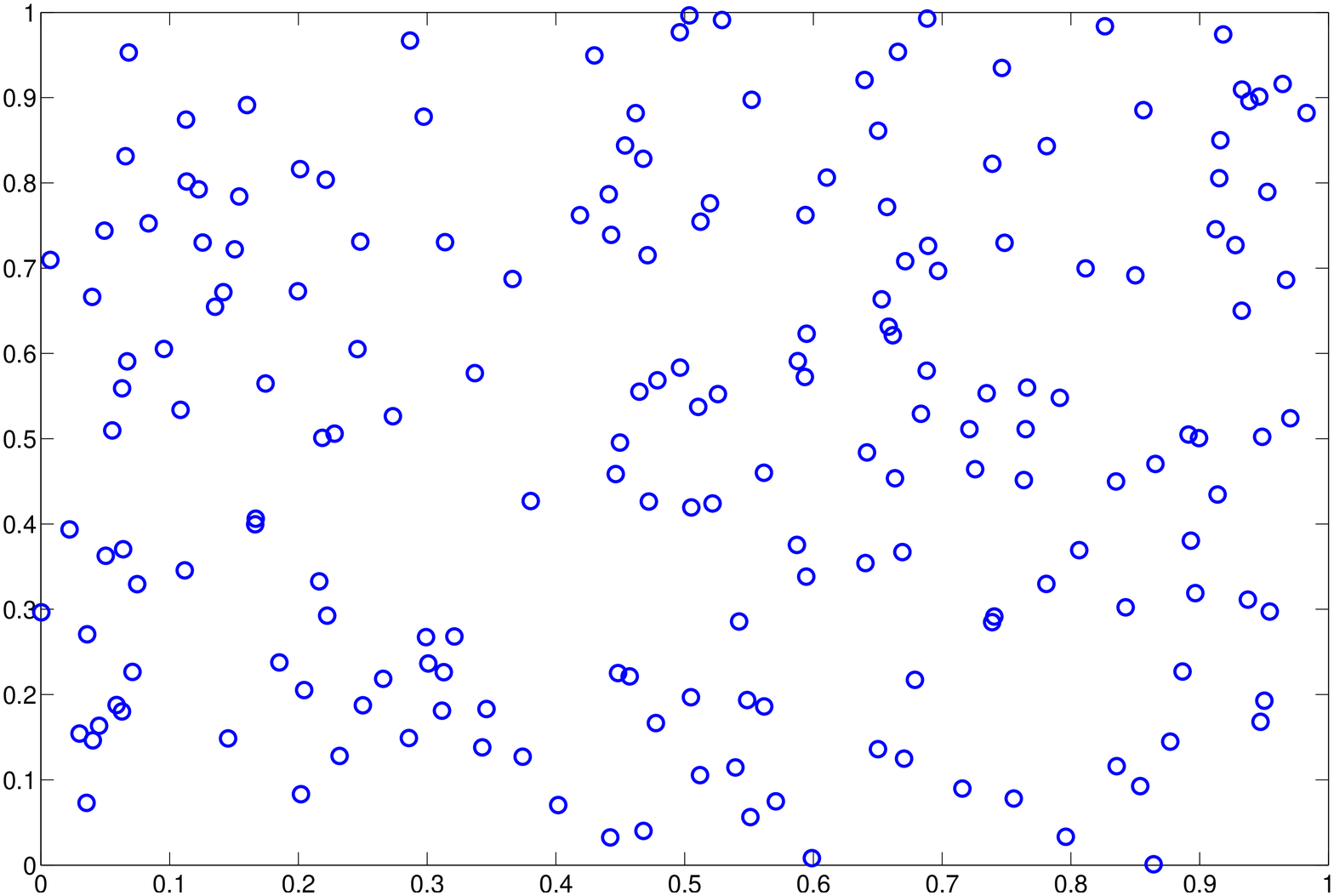}\ec
 \end{minipage}}\hfil\subfloat[b][Clustered $a=5$.]{\begin{minipage}{1in}
 \bc\includegraphics[width=1in,height=1in]{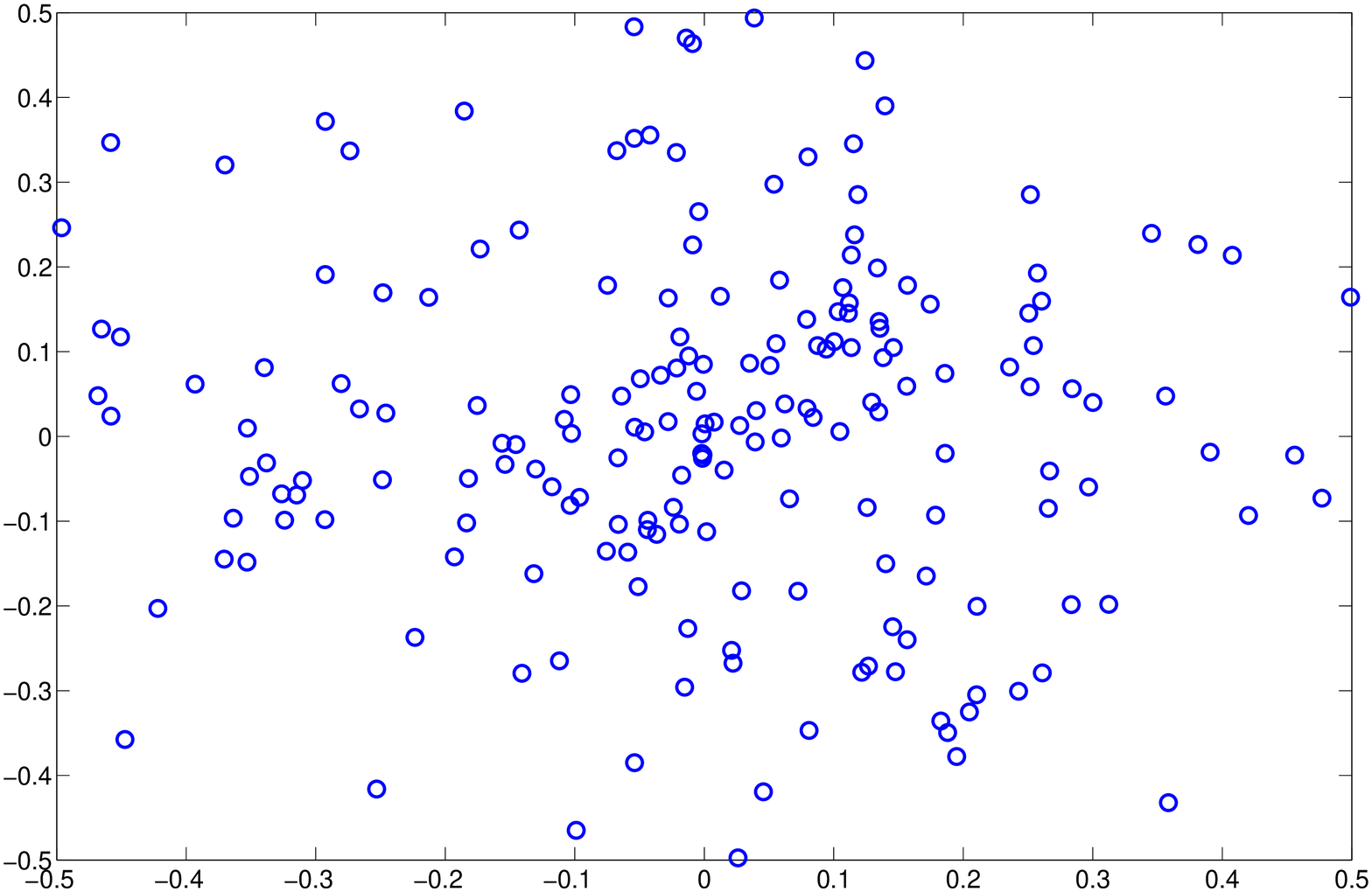}\ec
 \end{minipage}}\hfil\subfloat[c][Spread-out $a=-5$.]{\begin{minipage}{1.1in}
 \bc\includegraphics[width=1in,height=1in]{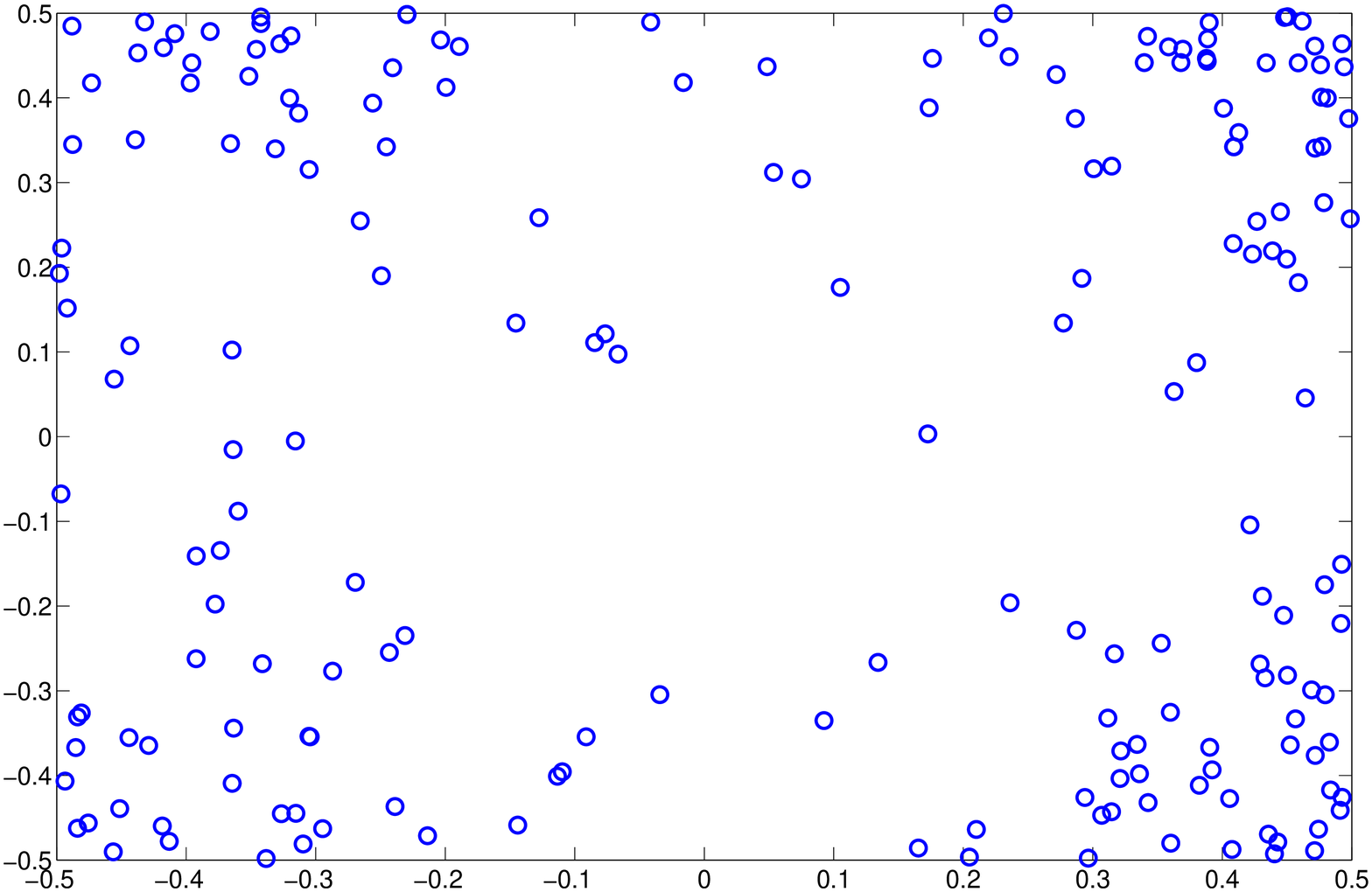}\ec
 \end{minipage}}
 \caption{Sample realization of $n=190$ points on unit square.
 See (\ref{eqn:kappa_symm}),
 (\ref{eqn:kappa_exp}).}\label{fig:points_kappa}
 \end{figure}

\begin{figure}[ht]
\begin{center}
 \begin{psfrags}
\psfrag{Average energy}[c]{\scriptsize Avg. energy for SPR} 
\psfrag{path loss nu}[c]{\scriptsize Path-loss exponent $\nu$} 
\psfrag{Uniform: a=0}[l]{\scriptsize Uniform: 
$a=0$}\psfrag{Clustered: a=10}[l]{\scriptsize Clustered: $a=5$} 
\psfrag{Spread out: a=-10}[l]{\scriptsize Spread out: $a=-5$}
\includegraphics[width=2.1in,height=1.6in]{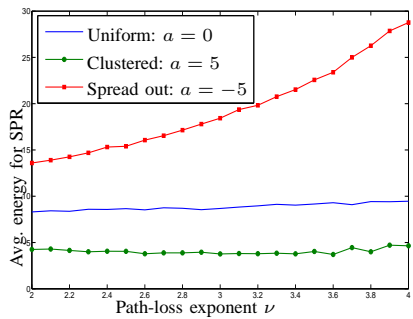}
\end{psfrags}
\end{center}\caption{Average energy for shortest-path routing policy over 500 
runs for node-placement pdfs shown in Fig.\ref{fig:points_kappa} and 
number of nodes  $n=190$.}\label{fig:spt_nu} \end{figure}

We now study the effect of i.i.d. node-placement pdf $\tau$ on the 
energy consumption of both DFMRF policy and shortest-path policy 
with no data aggregation. In Fig.\ref{fig:mst_kappa}, 
Fig.\ref{fig:disc_kappa_nu4} and Fig.\ref{fig:spt_nu}, we consider a 
family of truncated-exponential pdfs $\tau_a$ given by 
\beq\label{eqn:kappa_symm}\tau_a(x)= \xi_a(x(1)) \xi_a(x(2)),\quad 
x\in \R^2,\eeq where, for some $a\!\neq \!0$, $\xi_a$ is given by 
the truncated exponential\bcase{ \label{eqn:kappa_exp}\xi_a(z) 
\defeq}\nn \frac{a e^{-a |z|}}{2(1-e^{-\frac{a}{2}})},&if $z \in
[-\frac{1}{2},\frac{1}{2}]$,\\
0,&o.w.\ecase Note that as $a\!\!\to\!\! 0$, we obtain the uniform 
distribution in the limit  $(\tau_0 \equiv 1)$. A positive 
  $a$ corresponds to clustering   of the points  with 
respect to the origin and viceversa.
 In Fig.\ref{fig:points_kappa}, a sample
realization  is shown for the cases $a=\pm5$ and $a\! \to \! 0$.

Intuitively, for shortest-path   (SP) policy where there is no data 
aggregation, the influence of node placement on the energy 
consumption is fairly straightforward.  If we cluster the nodes 
close to one another,  the average energy consumption decreases. On 
the other hand, spreading the nodes out towards the boundary 
increases  the average energy. Indeed, we observe this behavior in 
Fig.\ref{fig:spt_nu}, for the placement pdf $\tau_a$ defined above 
in (\ref{eqn:kappa_symm}) and (\ref{eqn:kappa_exp}). However, as 
established in the previous sections, optimal node placement for the 
DFMRF policy does not follow this simple intuition.

%The behavior of the DFMRF policy under different node placements, is
%however, not so straightforward. Recall that the asymptotic bound
%for average energy  of DFMRF in
%(\ref{eqn:dfnngscalingclique_notscale}) comprises of  two terms, one
%corresponding to edges of the dependency graph, and the other, to
%the edges of the MST. They may behave differently for different
%placement pdfs $\tau$  depending on the dependency graph model and
%path loss $\nu$.

In  Theorem \ref{lemma:unif_mst}, we established that the uniform 
node placement $(\tau_0\equiv 1)$ minimizes the asymptotic average 
energy consumption of the optimal policy (which turns out to be the 
DFMRF policy), when  the path-loss exponent $\nu \geq 2$. For 
$\nu\in [0,2]$, the uniform distribution has the worst-case value. 
This is verified in Fig.\ref{fig:mst_kappa}, where  for $\nu \in 
[1,3]$, the uniform distribution initially has high energy 
consumption but decreases as we  increase the path-loss exponent 
$\nu$. We see that at threshold of around $\nu=2.4$, the uniform 
distribution starts having lower energy than the   non-uniform 
placements (clustered and spread-out), while according to Theorem 
\ref{lemma:unif_mst}, the   threshold should be $\nu=2$. Moreover, 
Theorem \ref{lemma:unif_mst}  also establishes that the clustered 
and spread-out distributions $(a\pm 5)$  have the same energy 
consumption since the expressions $\int_{\Bc_1} 
\tau_a(x)^{1-\frac{\nu}{2}}dx$ for $a=5$ and $a=-5$  are equal  for 
$\tau_a$ given by (\ref{eqn:kappa_symm}) and (\ref{eqn:kappa_exp}), 
and this approximately holds in Fig.\ref{fig:mst_kappa}.

%For The $K$-NNG Dependency Graph, From Theorem
%\Ref{Lemma:Uniform_Knng},   Uniform Node Placement Minimizes The
%Asymptotic Upper Bound On Average Energy. In 
%Fig.\Ref{Fig:Knng_Kappa_Nu}, For The $K$-NNG Dependency Graph, We 
%Plot The Ratio Of Energy Of DFMRF Under  Non-Uniform Placement With 
%Respect To The Energy Under 
%  Uniform Placement, And Also The Theoretical Value Of This 
%Ratio, Given By Corollary \Ref{Cor:Dfmrf} As
%\[\Int_{\Bc_1} \Kappa(X)^{1-\Frac{\Nu}{2}}Dx.\] For
%$\Kappa$ Given By (\Ref{Eqn:Kappa_Symm}) And (\Ref{Eqn:Kappa_Exp}), 
%And Find That The Above Expression Is Equal For $A=5$ And $A=-5$. We 
%Observe That The Simulation Results Are Close To The Theoretically 
%Predicted Value.

%In 
%Fig.\ref{fig:disk_nu}, we plot the average energy for DFMRF  for the 
%disc dependency graph with fixed disc radius $\delta=0.2$. We find 
%that for low values of $\nu \approx 2$, clustering nodes $(a=5)$ has 
%lower energy consumption than the uniform node placement, while the 
%situation is reversed as $\nu$ increases: the energy under uniform 
%placement is almost constant whereas the energy under non-uniform 
%placements increases rapidly. Spreading out the nodes $(a=-5)$ 
%performs worse than clustering throughout.
%
%
%In Fig.\ref{fig:disc_kappa_nu2}, we plot the average energy for
%DFMRF for path loss $\nu=2$. We find that for low values of the disc
%radius $\delta$, clustering $(a=5)$ performs better than   spreading
%out $(a=-5)$, which in turn is better than uniform placement. The
%situation is fully reversed at high radius $\delta$. 

 We now study the energy consumption of 
the DFMRF policy in Fig.\ref{fig:disc_kappa_nu4} under the disc 
dependency graph and  the   node placements given in 
Fig.\ref{fig:points_kappa}. In Fig.\ref{fig:disc_kappa_nu4},  for 
path-loss exponent $\nu=4$, we find that the uniform node placement  
$(\tau_0\equiv 1)$ performs significantly better than the 
non-uniform placements for the entire range of the disc radius 
$\delta$. Intuitively, this is because at large path-loss exponent 
$\nu$,  communication over long edges consumes a lot of energy and 
long edges occur with higher probability in non-uniform placements 
(both clustered and spread-out)  compared to the uniform placement. 
Hence, uniform node placement is significantly energy-efficient 
under high path-loss exponent of communication. 

%
%
%
%Intuitively, this is because the disc radius $\delta$ is relatively 
%small, and clustering the nodes does not add too many new edges to 
%the disc dependency graph.   The behavior is however reversed at 
%high values of disc radius $\delta$. In this case, clustering leads 
%to significant addition of new edges in the disc graph and increase 
%in the total costs. On the other hand, when nodes are spread out, 
%the addition of new edges is fewer at high $\delta$. 
%Hence, uniform placement has good performance under large path loss. 
%For a disc graph with higher $\delta$, clustering the nodes may add 
%significant number of new edges and hence, it is preferable to 
%spread out the nodes in this scenario.

%\subsection{Implications}

\section{Conclusion}\label{sec:conclusion}

We analyzed the scaling laws for energy consumption of data-fusion 
policies under the constraint of optimal statistical inference at 
the fusion center. Forwarding all the raw data without fusion has an 
unbounded average energy as we increase the network size, and hence, 
is not a feasible strategy in energy-constrained networks.  We 
established finite average energy scaling for a fusion policy known 
as the  Data Fusion for Markov Random Fields (DFMRF) for a class of 
spatial correlation model. We analyzed the influence of the 
correlation structure given by the dependency graph, the node 
placement distribution  and the transmission environment (path-loss 
exponent) on the energy consumption.

There are many issues which are not handled in this paper.  Our 
fusion policy DFMRF needs centralized network information for 
constructed, and we plan to investigate   distributed policies when 
only local  information is available at the nodes.  Our model 
currently only incorporates i.i.d. node placements and we expect our 
results to extend to the correlated node placement according to a 
Gibbs point process through the results in 
\cite{Schreiber&Yukich:08}. We have not considered here   the 
scaling behavior of the inference accuracy (error probability) with 
network size, and this is a topic of study in 
\cite{Anandkumar&Tong&Swami:09IT,Anandkumar&etal:09ISIT}.  We have 
not considered the time required for data fusion, and it is 
interesting to establish   bounds in this case. Our current 
correlation model assumes a discrete Markov random field. A more 
natural but difficult approach is to consider Markov field over a
continuous space \cite{Kunsch:79} and then, sample it through node 
placements.
%tradeoff....

%%Continuous MRF : better explanation for physical phenomena

\subsection*{Acknowledgment} \noindent The authors thank  
A.  Ephremides, T. He, D. Shah, the guest editor M. Haenggi and the 
anonymous reviewers  for helpful comments.

\begin{appendix}%
%\subsection{Proof of Lemma \ref{lemma:disc}}\label{proof:disc}
%
%Assume that dependency graph is the sub-critical disc graph or the
%continuum percolation graph on $\P_\tau$, where $\tau$ is a fixed
%scalar incident on $[ \inf_{x \in B_{1}} \tau(x), \sup_{x \in
%B_{1}} \tau(x) ]$ and let $\eta$ be defined as in (\ref{eqn:eta}).
%Given $\tau$ fixed, we argue that $\eta$ is  stabilizing on
%$\P_\tau$ as follows. Let $x \in \R^2$.
%
%Clearly the edges in $E_x$ have a length bounded by the diameter
%$D_x$ of the connected cluster belonging to the continuum
%percolation graph and which contains the clique containing $x$.  The
%radius of stabilization for $\eta$ at $x$ is bounded by the diameter
%$D_x$  since $\eta(x, \Pc_1)$ does not depend on point
%configurations outside the connected cluster at $x.$ If $\Pc_1$ is
%Poisson  and if the disc radius $\delta$ is subcritical, then the
%diameter $D_x$ of the connected cluster emanating from a given node
%$x$ has exponentially decaying tails, see Section 12.10 in
%\cite{Grimm:book}. Since $D_x$ is a.s. finite, this yields the
%required stabilization for $\eta$.
%
% Clearly $\eta$ satisfies the bounded moments condition
%(\ref{eqn:moment}) since each $R_{x,y}$ is bounded by the
%deterministic subcritical disc radius  and the number  of nodes in
%$n^{\frac{1}{2}} \{X_i\}_{i=1}^n$ which are joined to
%$n^{\frac{1}{2}} X_1$ is a random variable with moments of all
%orders.\qed

\subsection{Functionals on random points sets}\label{sec:lln}
 In
\cite{Penrose&Yukich:01AAP,Penrose&Yukich:02AAP,Penrose&Yukich:03AAP},
Penrose and Yukich introduce the concept of stabilizing functionals
to establish weak laws of large numbers for functionals on graphs
with random vertex sets.  As in this paper, the vertex sets may be
marked (sensor measurements constituting one example of marks), but 
for simplicity of exposition we  work with unmarked vertices.   We 
briefly describe the general weak law of large numbers after 
introducing the necessary definitions.

Graph functionals   on a vertex set $\Vmsc$ are often represented as
sums of spatially dependent terms  $$ \sum_{x \in \Vmsc}
\xi(x,\Vmsc),
$$ where $\Vmsc \subset \R^2$ is locally finite (contains only finitely many points in
any bounded region), and   the measurable function $\xi$, defined on 
all pairs $(x, \Vmsc)$, with $x \in \Vmsc$, represents the 
interaction of $x$ with  other points in $\Vmsc$. We see that the 
functionals corresponding to energy consumption  can be cast in this 
framework.

When $\Vmsc$ is random, the range of spatial dependence of $\xi$ at 
node $x \in \Vmsc$ is random, and the purpose of {\em stabilization 
} is to quantify this range in a way useful for asymptotic analysis.  
There are several similar notions of stabilization, but the essence 
is captured by the notion of stabilization of $\xi$ with respect to 
homogeneous Poisson points on $\R^2$, defined as follows. Recall 
that $\Pc_a$ is a homogeneous Poisson point process with intensity 
$a>0$.

We say that $\xi$ is translation invariant if $\xi(x,\Vmsc) = \xi(x
+ z, \Vmsc + z)$ for all $z \in \R^2$. Let $\0$ denote  the origin
of $\R^2$ and let $B_r(x)$ denote the Euclidean ball centered at $x$
with radius $r$. A translation-invariant $\xi$ is {\em homogeneously 
stabilizing} if for all intensities $a > 0$ there exists  almost 
surely a finite random variable $R:=R(a)$ such that $$ \xi(\0, 
(\Pc_{a} \cap B_R(\0) ) \cup \A ) = \xi(\0, \Pc_{a} \cap B_R(\0))$$ 
for all locally finite $\A \subset \R^2 \setminus B_R(\0)$. Thus 
$\xi$ stabilizes  if the value of $\xi$ at $\0$ is unaffected by 
changes in point configurations outside $B_R(\0)$.

%It is also useful to have a notion of stabilization for functionals
%of graphs.  Given a graph $\Gmsc$ and a node $x \in \Vmsc$, let
%$\Lmsc(x, \Gmsc(\Vmsc))$ be the set of links or edges of
%$\Gmsc(\Vmsc)$ incident on $x$. We shall say that $\Gmsc$ stabilizes
%on $\Pc_\tau$ if there exists a random but almost surely finite
%variable $R$ such that
%\[\Lmsc(\0;\Gmsc(\Pc_{\tau}\cup\{\0\})) = \Lmsc(\0;\Gmsc(\Pc_{\tau}\cup\{\0\}\cap
%B(\0;R))\cup \Ac),\] for all finite $\Ac \subset \R^2 \setminus
%B(\0;R)$. In other words, for a stabilizing graph, the edges of the
%origin are unaffected by the addition or deletion of points beyond
%some finite (random) distance.

$\xi$ satisfies the moment condition of order $p > 0$ if \beq  
\label{eqn:moment}
 \sup_{n \in \N} \E[ \xi(n^{\frac{1}{2}}X_1, n^{\frac{1}{2}}\{X_i \}_{i=1}^n)^p ] < \infty.
\eeq

We  use the following weak laws of large numbers throughout. Recall 
that $X_i$ are i.i.d. with density $\tau$.

\bt[WLLN \cite{Penrose&Yukich:07Bern,Penrose&Yukich:03AAP}] Put $q =
1$ or $q = 2$. Let $\xi$ be a homogeneously stabilizing  
translation-invariant functional satisfying the moment condition 
(\ref{eqn:moment}) for some $p > q$. Then
 \beqn\nn
&&\lim_{n \to \infty}  \frac{1}{n} \sum_{i=1}^n
\xi\Bigl(\sqrt{\frac{n}{\lambda}}X_i, \sqrt{\frac{n}{\lambda}} 
\{X_j\}_{j=1}^n \Bigr)
\\&=&
  \int_{\Bc_{1}}
\E [\xi(\0, \Pc_{\lambda\tau(x)})] \tau(x) dx  \ \ \text{in} \ \
L^q. \ \label{eqn:WLLN} \eeqn \et

\begin{figure}[tb]\bc\bp\psfrag{n to
infinity}[l]{\scriptsize $n \to 
\infty$}\psfrag{Origin}[l]{\scriptsize Origin} \psfrag{Normalized 
sum of edge weights}[l]{\scriptsize Normalized sum of 
edges}\psfrag{Expectation of edges}[l]{\scriptsize Expectation of 
edges}\psfrag{of origin of Poisson process}[l]{\scriptsize of origin 
of Poisson process}\psfrag{edge sum}[l]{\scriptsize $\frac{1}{n} 
\sum\limits_{e \in \Gmsc(\Vmsc_n)} |e|^\nu$}\psfrag{edge sum 
poisson}[l]{\scriptsize  
$\frac{1}{2}\lambda^{-\frac{\nu}{2}}\Ebb\!\! 
\!\!\!\!\sum\limits_{e\in \Lmsc(\0, \Gmsc(\Pc_{\lambda}\cup\{\0\}))} 
\!\!\!\!\!\! |e|^\nu $}\includegraphics[width = 
2.4in]{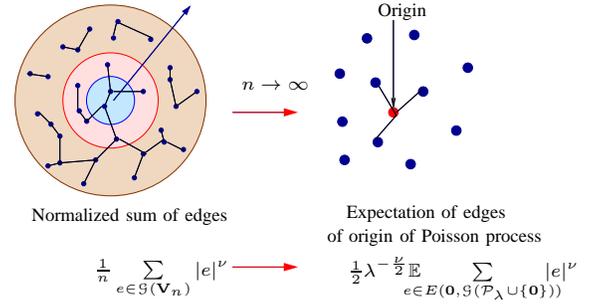}\ep\ec\caption{LLN for sum graph 
edges on  uniform point sets 
$(\tau\equiv1)$.}\label{fig:lln}\end{figure}

We interpret the right-hand side of the above equation as a weighted 
average of the values of $\xi$ on homogeneous Poisson point 
processes $\Pc_{\lambda \tau(x)}$. When $\xi$ satisfies scaling such 
as $\E[ \xi(\0, \Pc_{a})] = a^{-\alpha } \E [\xi(\0, \Pc_{1})]$,   
then the limit on the right-hand side of (\ref{eqn:WLLN}) simplifies 
to \beq\label{eqn:homo} \la^{-\alpha} \E [\xi(\0, \Pc_{1})] 
\int_{\Bc_{1}} (\tau(x))^{1- \alpha}dx \ \ \text{in} \ \ L^q,\eeq a 
limit appearing regularly in problems in Euclidean combinatorial 
optimization. For uniform node placement $(\tau(x)\equiv 1)$, the 
expression in (\ref{eqn:WLLN}) reduces to $ \E [\xi(\0, 
\Pc_{\lambda})]$, and the LLN result for this instance is 
pictorially depicted in Fig.\ref{fig:lln}.

For example, if $\xi(x, \Vmsc)$ is one half the sum of the 
$\nu$-power weighted edges incident to $x$ in the MST (or any 
scale-invariant stabilizing graph) on $\Vmsc$, i.e.,
$$
\xi(x, \Vmsc)\defeq \frac{1}{2}  \sum_{e \in E(x,\smst(\Vmsc))} 
|e|^\nu,
$$
then substituting $\alpha$ with $\frac{\nu}{2}$ in (\ref{eqn:homo}), 
\beqn\nn &&\lim_{n\to\infty} \frac{1}{n} \sum_{i=1}^n 
\xi\Bigl(\sqrt{\frac{n}{\lambda}}X_i, \sqrt{\frac{n}{\lambda}} \{X_i 
\}_{i=1}^n\Bigr)
\\ \nn
 &=& \lambda^{-\frac{\nu}{2}} \E [\xi(\0, \P_1)] \int_{Q_1} (\tau(x))^{1 - 
\frac{\nu}{2}} dx\\   &=&  \lambda^{-\frac{\nu}{2}} \zeta(\nu; \mst) 
\int_{Q_1} (\tau(x))^{1 - \frac{\nu}{2}} 
dx,\label{eqn:constgd_derivation}\eeqn where $\zeta(\nu;\mst)$ is 
defined in (\ref{eqn:constgd_def_mst}).

\subsection{Proof of Theorem \ref{thm:dfmrf_scaling}}\label{proof:dfmrf_scaling}

 The energy consumption of DFMRF satisfies the inequality in
(\ref{eqn:lcg_lag_bound_norm}). For the MST we have the result in
Theorem \ref{thm:iid_scaling}. We now use stabilizing functionals to 
show that \[\frac{1}{n} \sum_{e \in \Gmsc(\Vmsc_n)}|e|^\nu\] 
converges in $L^2$ to a constant.   For all locally finite vertex 
sets ${\cal X} \subset \R^2$ supporting some dependency graph 
$\Gmsc(\Xc)$ and for all $x \in {\cal X}$, define the functional 
$\eta(x, {\cal X})$ by \beq \label{eqn:eta}\eta(x, {\cal X})
\defeq   \sum_{y: (x,y)\in \sdep(\Xc)}|x,y|^\nu.\eeq % each edge counted twice since i<j is not imposed.
 Notice that $\sum_{x \in {\cal X}} \eta(x, {\cal X}) =2 \sum_{e
\in \sdep({\cal X})}  |e|^\nu$.

 From \cite[Thm 2.4]{Penrose&Yukich:03AAP}, the sum of power-weighted edges  of the 
 $k$-nearest neighbors graph is a  stabilizing functional and satisfies
the bounded-moments condition (\ref{eqn:moment}). Hence, the limit
in (\ref{eqn:WLLN}) holds when the dependency graph is the
$k$-NNG.

Finally,  the  sum of power-weighted edges  of the continuum 
percolation graph is a  stabilizing functional which satisfies the 
bounded-moments condition (\ref{eqn:moment}), thus implying that the 
limit in (\ref{eqn:WLLN}) holds.

Indeed, $\eta$ stabilizes with respect to  $\P_a$, $ a \in (0, 
\infty)$, since  points distant from $x$ by more than the 
deterministic disc radius do not modify the value of $\eta(x, 
\P_a)$.  Moreover, $\eta$ satisfies the bounded moments condition 
(\ref{eqn:moment}) since each $|x,y|$ is bounded by the 
deterministic disc radius  and the number of nodes in 
$n^{\frac{1}{2}} \{X_i\}_{i=1}^n$ which are joined to 
$n^{\frac{1}{2}} X_1$ is a random variable with moments of all 
orders.\qed

%Assume that the dependency graph is the sub-critical disc graph
%(continuum percolation) on $\P_a$, where $\tau$ is a fixed scalar
%belonging to $[ \inf_{x \in \Bc_{1}} \tau(x), \sup_{x \in \Bc_{1}}
%\tau(x) ]$ and let $\eta$ be defined as in (\ref{eqn:eta}).  Given
%$\tau$ fixed, we argue that $\eta$ is stabilizing on $\P_\tau$ as
%follows. Let $x \in \R^2$.

%Clearly the edges in $E_x$ have a length bounded by the diameter
%$D_x$ of the connected cluster belonging to the continuum
%percolation graph and which contains the clique containing $x$.  The
%radius of stabilization for $\eta$ at $x$ is bounded by the diameter
%$D_x$  since $\eta(x, \Pc_1)$ does not depend on point
%configurations outside the connected cluster at $x.$ If the disc
%radius $\delta$ is subcritical, then the diameter $D_x$ of the
%connected cluster emanating from a given node $x$ has
%exponentially decaying tails, see Section 12.10 in
%\cite{Grimm:book}. Since $D_x$ is a.s. finite, this yields the
%required stabilization for $\eta$.

%
%By Lemma \ref{lemma:knng} and \ref{lemma:disc},  the functional
%$\eta$ is stabilizing and satisfies the bounded-moments condition in
%(\ref{eqn:moment}) when the dependency graph is either the $k$-NNG
%or the sub-critical continuum percolation graph.  Hence, the limit
%in (\ref{eqn:WLLN}) holds. \qed
\end{appendix}

{\scriptsize
% Generated by IEEEtran.bst, version: 1.12 (2007/01/11)

%\bibliographystyle{IEEETran}
%\bibliography{IEEEabrv,\bibhome/Journal,\bibhome/Conf,\bibhome/Misc,\bibhome/ACSP-J,\bibhome/Book}
}

\begin{biography}[{\includegraphics[width=1in,height
=1.25in,clip,keepaspectratio]{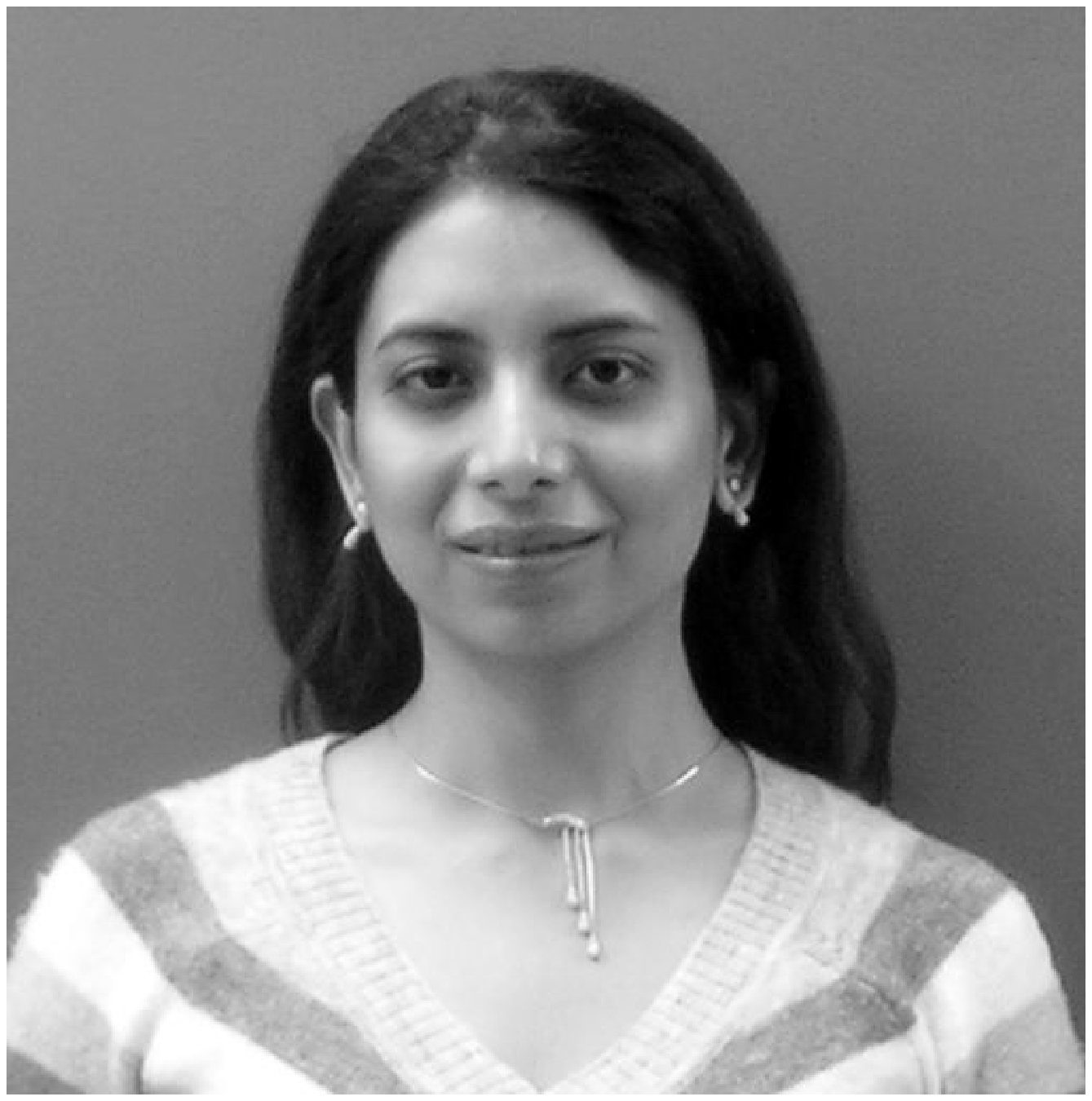}}]{\bf 
Animashree Anandkumar} (S `02) received her B.Tech in Electrical 
Engineering from the Indian Institute of Technology Madras in 2004 
with a minor in Theoretical Computer Science. She is a PhD student 
in Electrical Engineering at Cornell University with a minor in 
Applied Mathematics. Since Fall 2008, she is visiting the Stochastic 
Systems Group at MIT, Cambridge, MA.

Anima received the 2008 IEEE Signal Processing Society (SPS) Young 
Author award for her paper co-authored with Lang Tong appearing in 
the IEEE Transactions on Signal Processing. She is the recipient of 
the Fran Allen IBM Ph.D fellowship for the year 2008-09,  presented 
annually to one female Ph.D. student in conjunction with the IBM 
Ph.D. Fellowship Award. She  was named a finalist for the Google 
Anita-Borg Scholarship 2007-08.  She received the Student Paper 
Award  at the 2006 International Conference on Acoustic, Speech and 
Signal Processing  (ICASSP).   Her research interests are in the 
area of statistical-signal processing, information theory and 
networking. Specifically, she has been worked on distributed 
inference and learning of graphical models, routing and 
random-access schemes, error exponents and queueing models. She has 
served as a reviewer for IEEE Transactions on Signal Processing, 
IEEE Transactions on Information Theory, IEEE Transactions on 
Wireless Communications and IEEE Signal Processing Letters.
\end{biography}

\begin{biography}[{\includegraphics[width=1in,height
=1.25in,clip,keepaspectratio]{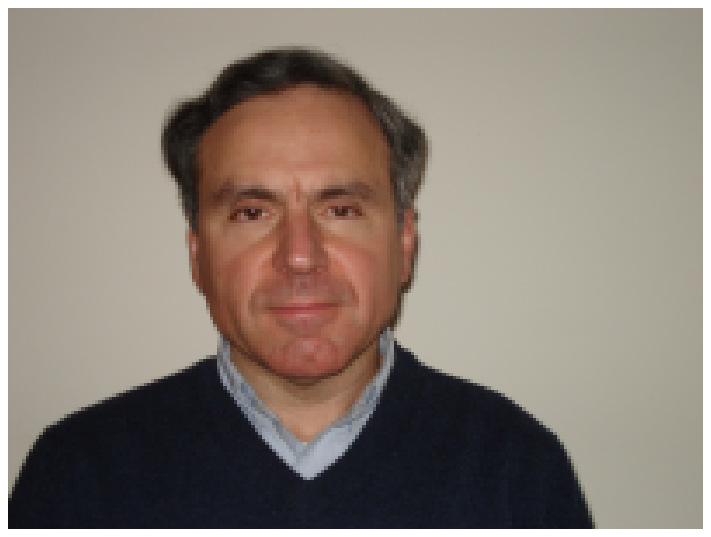}}]{\bf Joseph E. 
Yukich} received the B.A. and Ph.D. degrees in mathematics from 
Oberlin College (1978) and the Massachusetts Institute of Technology 
(1983), respectively.   He is the recipient of two Fulbright awards 
to France and has held visiting positions in Australia, Bell Labs, 
Cornell and Switzerland.  He  is currently a member of the faculty 
at Lehigh University.

He has research interests in spatial point processes, stochastic 
geometry,  combinatorial optimization, and random networks  and has 
authored the monograph Probability Theory of Classical Euclidean 
Optimization Problems,  (Lecture Notes in Mathematics, volume 1675). 
His research has been supported by the National Science Foundation 
and the National Security Agency. \end{biography}

\begin{biography}[{\includegraphics[width=1in,height
=1.25in,clip,keepaspectratio]{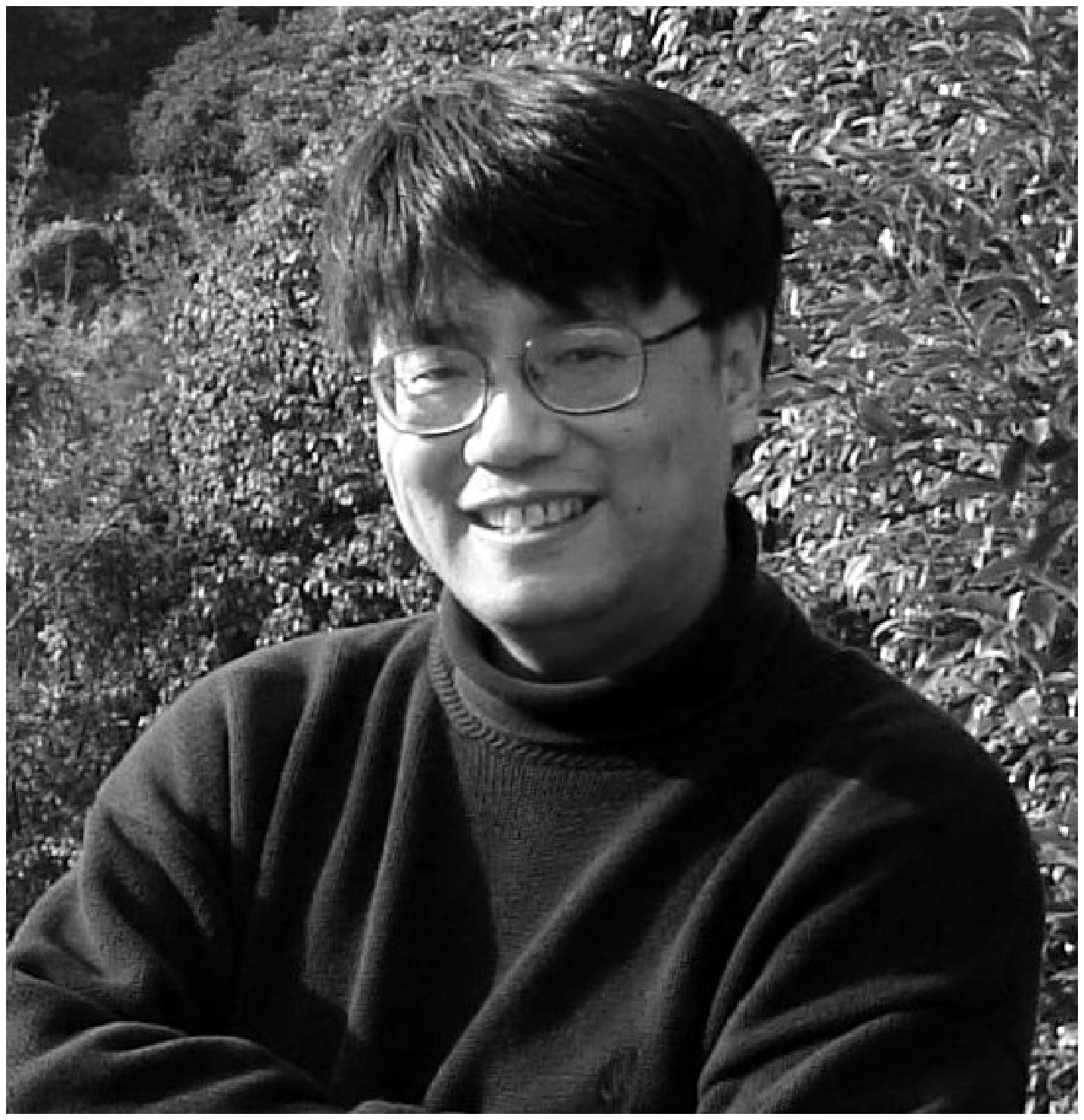}}]{\bf Lang Tong} 
(S'87,M'91,SM'01,F'05) is the Irwin and Joan Jacobs Professor in 
Engineering at Cornell University Ithaca, New York.  Prior to 
joining Cornell University, he was on faculty at the West Virginia 
University and the University of Connecticut. He was also the 2001 
Cor Wit Visiting Professor at the Delft University of Technology. 
Lang Tong received the B.E. degree from Tsinghua University, 
Beijing, China, in 1985, and M.S. and Ph.D. degrees in electrical 
engineering in 1987 and 1991, respectively, from the University of 
Notre Dame, Notre Dame, Indiana. He was a Postdoctoral Research 
Affiliate at the Information Systems Laboratory, Stanford University 
in 1991.

Lang Tong is a Fellow of IEEE. He received the 1993 Outstanding 
Young Author Award from the IEEE Circuits and Systems Society, the 
2004 best paper award (with Min Dong) from IEEE Signal Processing 
Society, and the 2004 Leonard G. Abraham Prize Paper Award from the 
IEEE Communications Society (with Parvathinathan Venkitasubramaniam 
and  Srihari Adireddy). He is also a coauthor of five student paper 
awards. He received Young Investigator Award from the Office of 
Naval Research.  

Lang Tong's research is in the general area of statistical signal 
processing, wireless communications and networking, and information 
theory.  He has served as an Associate Editor for the IEEE 
Transactions on Signal Processing, the IEEE Transactions on 
Information Theory, and IEEE Signal Processing Letters.

\end{biography}

\begin{biography}[{\includegraphics[width=1in,height 
=1.25in,clip,keepaspectratio]{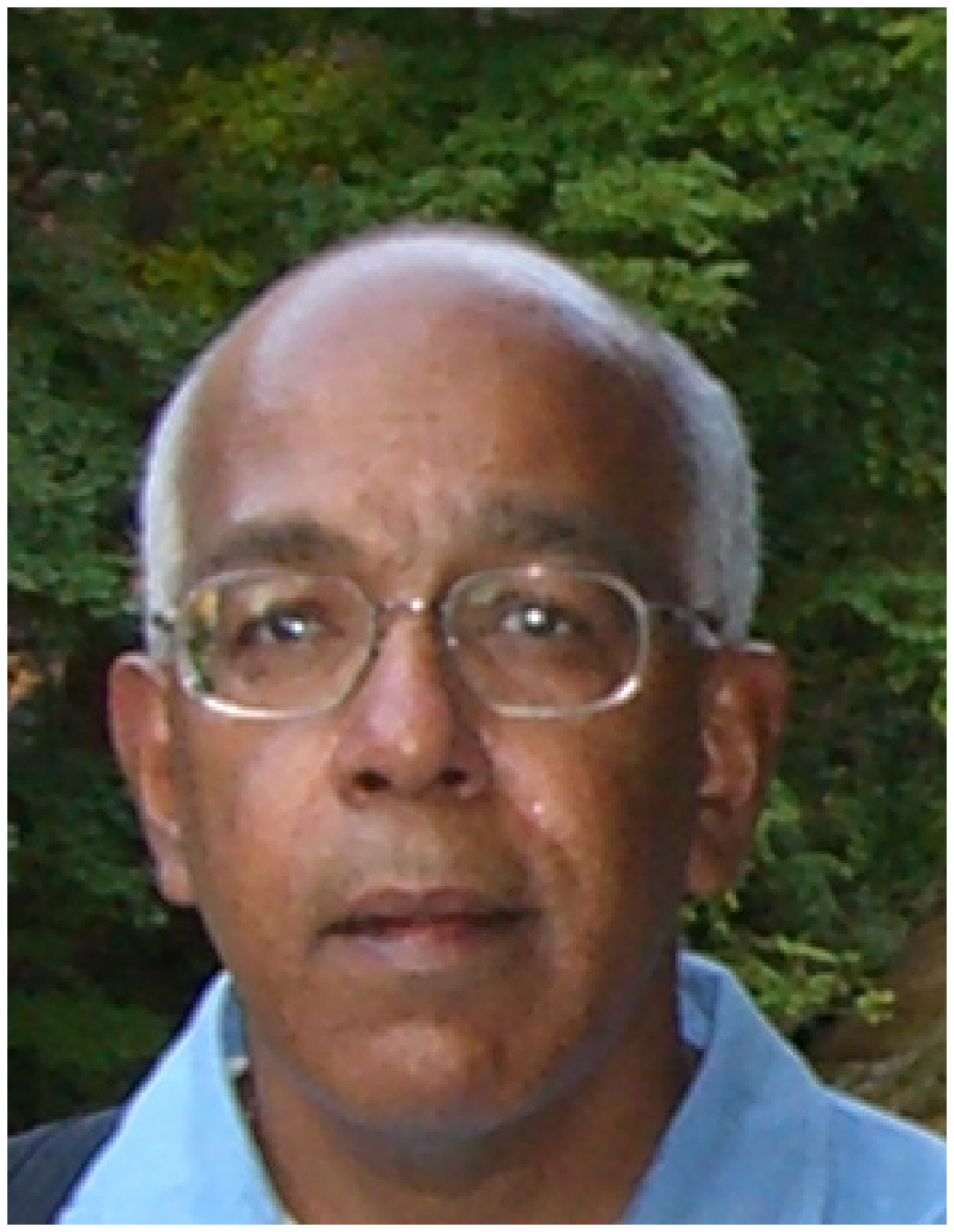}}]{\bf Ananthram 
Swami}   received the B.Tech. degree from IIT-Bombay; the M.S. 
degree from Rice University, and the Ph.D. degree from the 
University of Southern California (USC), all in Electrical 
Engineering.  He has held positions with Unocal Corporation, USC, 
CS-3 and Malgudi Systems.  He was a Statistical Consultant to the 
California Lottery, developed a Matlab-based toolbox for 
non-Gaussian signal processing, and has held visiting faculty 
positions at INP, Toulouse. He is with the US Army Research 
Laboratory (ARL) where his work is in the broad areas of signal 
processing, wireless communications, sensor and mobile ad hoc 
networks. He is an ARL Fellow.

Dr. Swami is a member of the IEEE Signal Processing Society's (SPS) 
Technical Committee (TC) on Sensor Array \& Multi-channel systems, a 
member of the IEEE SPS Board of Governors, and serves on the Senior 
Editorial Board of IEEE Journal of Selected Topics in Signal 
Processing. He has served as: chair of the IEEE SPS TC on Signal 
Processing for Communications; member of the SPS Statistical Signal 
and Array Processing TC; associate editor of the IEEE Transactions 
on Wireless Communications, IEEE Signal Processing Letters, IEEE 
Transactions on Circuits \& Systems-II, IEEE Signal Processing 
Magazine, and IEEE Transactions on Signal Processing. He was guest 
editor of a 2004 Special Issue (SI) of the IEEE Signal Processing 
Magazine (SPM) on `Signal Processing for Networking', a 2006 SPM SI 
on `Distributed signal processing in sensor networks', a 2006 
EURASIP JASP SI on Reliable Communications over Rapidly Time-Varying 
Channels', a 2007 EURASIP JWCN SI on `Wireless mobile ad hoc 
networks', and the Lead Editor for a 2008 IEEE JSTSP SI on ``Signal 
Processing and Networking for Dynamic Spectrum Access''.  He is a 
co-editor of the 2007 Wiley book ``Wireless Sensor Networks: Signal 
Processing \& Communications Perspectives''. He has co-organized and 
co-chaired three IEEE SPS Workshops (HOS'93, SSAP'96, SPAWC'10) and 
a 1999 ASA/IMA Workshop on Heavy-Tailed Phenomena.  He has co-led 
tutorials on `Networking Cognitive Radios for Dynamic Spectrum 
Access'' at ICASSP 2008, DySpan 2008 and MILCOM 2008.

\end{biography}

\end{document}